\documentclass{article}
\usepackage{emulateapj}
\usepackage{amsmath}
\usepackage{amssym}
\usepackage{psfig}
\usepackage{times}

\input{epsf}

\def\ph{\,{\rm ph}}

\def\Lobs{L_{\rm obs}}
\def\tobs{t_{\rm obs}}
\def\xiD{\xi_{\Delta}}

\def\trise{t_{\rm rise}}
\def\tpeak{t_{\rm peak}}

\def\tsw{t_{\rm sw}}
\def\Rdec{R_{\rm dec}}
\def\mdec{m_{\rm dec}}
\def\taugap{\tau_{\rm gap}}
\def\mgap{m_{\rm gap}}
\def\gdec{\gamma_{\rm dec}}
\def\xdec{x_{\rm dec}}
\def\xload{x_{\rm load}}
\def\fload{f_{\rm load}}
\def\facc{f_{\rm acc}}

\def\macc{m_{\rm acc}}

\def\Eacc{E_{\rm acc}}
\def\Rload{R_{\rm load}}
\def\Racc{R_{\rm acc}}
\def\Dej{\Delta_{\rm ej}}
\def\Gej{\Gamma_{\rm ej}}
\def\bej{\beta_{\rm ej}}
\def\Mej{M_{\rm ej}}
\def\Eej{E_{\rm ej}}
\def\gsat{\gamma_{\rm sat}}
\def\bsat{\beta_{\rm sat}}
\def\Rem{R_{\rm em}}
\def\Rsat{R_{\rm sat}}
\def\Rgap{R_{\rm gap}}
\def\xgap{x_{\rm gap}}
\def\l1{\lambda_1}
\def\ph{\hat{\phi}}
\def\dnp{\dot{n}_+}
\def\als{\alpha_1}
\def\alh{\alpha_2}
\def\dd{{\rm d}}
\def\dM{\dot{M}}
\def\sT{\sigma_{\rm T}}
\def\sKN{\sigma_{\rm KN}}
\def\FT{F_{\rm T}}
\def\XT{X^{\rm T}}

\def\taucr{\tau_{\rm cr}}

\def\tobs{t_{\rm obs}}

\def\ThC{\Theta_{\rm C}}
\def\gC{\gamma_{\rm C}}
\def\ge{\gamma_e}
\def\gsc{\gamma_{\rm sc}}
\def\gmax{\gamma_{\rm max}}
\def\gc{\gamma_c}
\def\Rmax{R_{\rm max}}
\def\Rgap{R_{\rm gap}}
\def\xiC{\xi_{\rm C}}
\def\xiload{\xi_{\rm load}}
\def\xiacc{\xi_{\rm acc}}
\def\tacc{t_{\rm acc}}
\def\xisc{\xi_{\rm sc}}
\def\xibr{\xi_{\rm br}}
\def\xisat{\xi_c}
\def\ximix{\xi_{\rm mix}}

\def\Erad{E_{\rm rad}}
\def\Ed{E_{\rm diss}}
\def\bh{\hat{\beta}}
\def\G0{\Gamma_0}
\def\bp{\beta^\prime}
\def\nup{\nu_s}
\def\mheat{m_{\rm heat}}
\def\tnup{\tilde{\nu}_s}
\def\vp{\varpi}
\def\vpp{\varpi^\prime}

\def\vpcr{\varpi_{\rm cr}}
\def\gcr{\gamma_{\rm cr}}
\def\vpload{a}
\def\vpacc{\varpi_{\rm acc}}
\def\vpbr{\varpi_{\rm br}}

\def\vpsc{\vp_{\rm sc}}
\def\ep{\epsilon}
\def\epsc{\epsilon_{\rm sc}}
\def\epthr{\epsilon_{\rm thr}}
\def\epabs{\epsilon_{\rm abs}}
\def\epKN{\epsilon_{\rm KN}}
\def\epbr{\epsilon_{\rm br}}
\def\epmin{\epsilon_{\rm min}}
\def\epmax{\epsilon_{\rm max}}
\def\ginj{\gamma_{\rm inj}}
\def\pinj{p_{\rm inj}}
\def\Isc{I_{\rm sc}}
\def\Psc{\dot{P}_{\rm sc}}
\def\Ppm{\dot{P}_\pm}
\def\lgg{\lambda_{\gamma\gamma}}
\def\tgg{\tau_{\gamma\gamma}}
\def\Rgg{R_{\gamma\gamma}}
\def\kgg{\kappa_{\gamma\gamma}}
\def\sgg{\sigma_{\gamma\gamma}}

\newbox\grsign \setbox\grsign=\hbox{$>$} \newdimen\grdimen \grdimen=\ht\grsign
\newbox\simlessbox \newbox\simgreatbox \newbox\simpropbox
\setbox\simgreatbox=\hbox{\raise.5ex\hbox{$>$}\llap
     {\lower.5ex\hbox{$\sim$}}}\ht1=\grdimen\dp1=0pt
\setbox\simlessbox=\hbox{\raise.5ex\hbox{$<$}\llap
     {\lower.5ex\hbox{$\sim$}}}\ht2=\grdimen\dp2=0pt
\setbox\simpropbox=\hbox{\raise.5ex\hbox{$\propto$}\llap
     {\lower.5ex\hbox{$\sim$}}}\ht2=\grdimen\dp2=0pt
\def\simgt{\mathrel{\copy\simgreatbox}}
\def\simlt{\mathrel{\copy\simlessbox}}

\newcommand{\bez}{\begin{eqnarray*}}
\newcommand{\eez}{\end{eqnarray*}}
\newcommand{\be}{\begin{equation}}
\newcommand{\ee}{\end{equation}}
\newcommand{\beq}{\begin{eqnarray}}
\newcommand{\eeq}{\end{eqnarray}}
\newcommand{\bc}{\begin{center}}
\newcommand{\ec}{\end{center}}

\lefthead{BELOBORODOV} 
\righthead{RADIATION FRONT OF GAMMA-RAY BURSTS}

\begin{document}

\title{Radiation front sweeping the ambient medium of gamma-ray bursts}

\author{Andrei M. Beloborodov\altaffilmark{1}} 
\affil{Stockholm Observatory, SCFAB, SE-106 91, Stockholm, Sweden;
andrei@astro.su.se} 
\altaffiltext{1}{Also at Astro-Space Center of Lebedev Physical Institute, 
Profsojuznaja 84/32, Moscow 117810, Russia}

\begin{abstract}

Gamma-ray bursts (GRBs) 
are emitted by relativistic ejecta from powerful cosmic explosions. 
Their light curves suggest that the $\gamma$-ray emission occurs at
early stages of the ejecta expansion, well before it decelerates in 
the ambient medium. If so, the launched $\gamma$-ray front must overtake 
the ejecta and sweep the ambient medium outward. 
As a result a gap is opened between the ejecta and the medium that surfs 
the radiation front ahead. Effectively, the ejecta moves in a cavity 
until it reaches a radius $\Rgap\approx 10^{16}E_{54}^{1/2}$cm where $E$ 
is the isotropic energy of the GRB. At $R=\Rgap$ the gap is closed, 
a blast wave forms and collects the medium swept by radiation. 
Further development of the blast wave is strongly affected by
the leading radiation front: the front plays the role of a precursor 
where the medium is loaded with $e^\pm$ pairs and preaccelerated just 
ahead of the blast. It impacts the emission from the blast 
at $R<\Rload=5\Rgap$ (the early afterglow). A spectacular observational 
effect results: {\it GRB afterglows should start in optical/UV 
and evolve fast ($<$~min) to a normal X-ray afterglow}. The early 
optical emission observed in GRB~990123 may be explained in this way.
The impact of the front is especially strong if the 
ambient medium is a wind from a massive progenitor of the GRB. 
In this case three phenomena are predicted:
(1) The ejecta decelerates at $R<\Rload$ producing
a lot of soft radiation. (2) The light curve of soft emission peaks at 
$\tpeak\approx 40(1+z)E_{54}^{1/2}(\Gej/100)^{-2}$s
where $\Gej$ is the Lorentz factor of the ejecta. 
Given measured redshift $z$ and $\tpeak$, one finds $\Gej$.
(3) The GRB acquires a spectral break at $5-50$~MeV because
harder photons are absorbed by radiation scattered in the wind.
A measurement of the break position will determine the wind density.

\end{abstract}


\section{Introduction}

Cosmological gamma-ray bursts (GRBs) are explosions of huge energy 
$\sim 10^{52}-10^{54}$~ergs (see Piran 1999 for a review). The relativistic 
ejecta 
of the explosion produces the observed $\gamma$-ray pulse with duration of 
a few seconds that propagates ahead of the ejecta and interacts with the 
ambient medium first, before the blast wave driven by the ejecta.

Madau \& Thompson (2000) and Thomson \& Madau (2000, hereafter TM) pointed 
out that the $\gamma$-ray pulse can preaccelerate the ambient medium to a 
high Lorentz factor. Even more importantly, the pulse-medium interaction 
is accompanied by runaway loading of $e^\pm$ pairs (TM). The interaction 
occurs inside the thin radiation front where
the primary photons scatter off the medium and 
turn into $e^\pm$ via $\gamma-\gamma$ reaction;
the created $e^\pm$ increase the medium opacity, do more scattering, and 
next generations of $e^\pm$ are loaded in a runaway manner. 
TM also pointed out that the $e^\pm$ loading and preacceleration ahead 
of the blast wave should modify the GRB afterglows 
(see also recent paper by M\'esz\'aros, Ramirez-Ruiz, \& Rees 2001). 

In the present paper we develop an accurate model for the radiation front
and assess its impact on the blast wave. The propagating front
is self-similar and its non-linear structure (the rise
of density and velocity across the front) admits a simple 
description. We find the medium parameters behind the front and identify
the range of radii where the impact on the medium is strong and hence the
ensuing blast wave is strongly affected. Especially interesting effects
are found for GRBs with massive progenitors, leading to spectacular
observational phenomena.

In this paper we assume that the radiation front is formed early 
inside the ejecta (the so-called ``internal'' scenario which
easily explains the fast variability of GRBs, see Piran 1999).
The front energy remains constant with radius 
i.e. we neglect additional radiation from the ensuing blast wave (afterglow) 
compared to the prompt radiation from the ejecta. In a separate paper
(Beloborodov A.M., in preparation) we study  
the opposite case where the ejecta emission is negligible
and the $\gamma$-ray front is created by the blast wave itself
(the ``external''~scenario).

In \S\S~2 and 3 a detailed formulation of the problem and 
basic equations are given.
Numerical solution is presented in \S~4. In \S~5 we develop
an analytical model that explains the front structure and reproduces 
the numerical results with good accuracy. The backreaction of
the GRB-medium interaction on the prompt $\gamma$-rays is studied in \S~6.
The sweeping of the medium by radiation and the front evolution
with radius are studied in \S~7. In \S~8 we compute the blast wave dynamics
in the preaccelerated environment and evaluate its emission.


\section{Formulation of the problem}

\subsection{Basic parameters of the front}

GRB produces a thin shell (``front'') of collimated radiation with 
bolometric flux $F(\vp)$ and spectrum $F_\ep(\vp)$ where $\ep=h\nu/m_ec^2$ 
and $\vp$ is the Lagrangian coordinate in the moving shell, 
$0<\vp<\Delta$. Here $\Delta$ is the front thickness and $\Delta/c$ 
is the observed duration of the burst. The front propagates through the 
ambient medium with velocity $c$. 
The medium interaction with the front is convenient to view in the 
$\vp$-coordinate:
medium ``enters'' the $\Delta$-shell at $\vp=0$, passes through it and goes out
at $\vp=\Delta$ with new density and velocity.
Radiation scattered by the medium is decollimated and 
also streams toward large $\vp$, being absorbed by the primary beam.

The scattering of GRB radiation can have a strong impact on the medium if 
each electron scatters many photons during its passage through the 
$\Delta$-shell. The photons ``kicked out'' by the electron from the 
collimated $\gamma$-ray beam can be converted into $e^\pm$, so a large 
number of scatterings would imply a large number of pairs created per one 
ambient electron. The main contribution to pair production comes from 
photons with $\ep\sim 1$ (see \S~5), and their density is 
$n_{\rm ph}\sim F/m_ec^3$. The electron scatters many photons if its 
``free path'' $\lambda=1/n_{\rm ph}\sT$ (the difference $\delta\vp$ between 
successive scatterings) is smaller than the front 
width,\footnote{Standard notation is used throughout the paper: 
a magnitude $Y$ measured in CGS units and divided by $10^k$ is denoted as 
$Y_k$.} 
\begin{equation}
\label{eq:lam}
 \lambda=\frac{m_ec^3}{F\sT}=4.64\times 10^6
                             R_{15}^2L_{53}^{-1}{\rm ~cm}<\Delta.
\end{equation}
The radiation flux is $F=L/4\pi R^2$
where $R$ is the distance from the center of the explosion
and $L$ is the isotropic luminosity of the GRB.
The total energy of the radiation
pulse is $E=(\Delta/c)L$ and the condition~(\ref{eq:lam}) can be rewritten as
\begin{equation}
\label{eq:Rlam}
 R<R_\lambda=\left(\frac{E\sT}{4\pi m_ec^2}\right)^{1/2}
  =8.0\times 10^{16}E_{53}^{1/2}{\rm ~cm}.
\end{equation}

Beside $\lambda$ there is another important length-scale in the front 
--- the typical $\delta\vp$ the scattered photons pass
before they get absorbed by the primary radiation. 
This ``photon free path'' (hereafter denoted $\lgg$) 
far exceeds $\lambda$. It implies that pair creation occurs at 
larger $\vp$ i.e. substantially lags behind scattering.
As we show in \S~5, the runway pair loading starts at 
$\vp=a\approx\sqrt{\lambda\lgg}\approx 30\lambda$.
Efficient pair loading in the front requires $\Delta>30\lambda$ which 
implies a tighter constraint on radius: $R<\Rload=R_\lambda/\sqrt{30}$.
At radii larger than $\Rload$ there is neither pair loading 
nor acceleration of the medium.

$\Rload$ should be compared with the deceleration radius of the GRB ejecta. 
The standard 
$R_{\rm dec}\sim 10^{16}-10^{17}$~cm $> \Rload$ holds if the ejecta 
decelerates in a normal interstellar medium (ISM). 
However, if the GRB has a massive progenitor, its ambient medium is 
the wind from the progenitor; then 
$R_{\rm dec}\sim 10^{14}-10^{16}$cm 
depends on the mass loss $\dM$ and the velocity $w$ of the wind, 
e.g. $R_{\rm dec}\sim 10^{15}$cm for $\dM=10^{-5}M_\odot~{\rm yr}^{-1}$ 
and $w=10^8$~cm~s$^{-1}$ expected for a Wolf-Rayet progenitor 
(Chevalier \& Li 1999). The impact of the radiation front is 
especially strong in the latter scenario where $\Rdec<\Rload$. 
In particular, as shown in \S~8, the very value of $\Rdec$ is changed.

The front model we construct makes use of the fact that
the scattering optical depth of the ambient medium $\tau_R=\sT R\rho_0/m_p$ 
is very small. The GRB pulse-medium interaction occurs
in a specific regime: each electron of the medium experiences a lot of 
scattering while GRB photons have a low probability of scattering.
(Pair loading increases the optical depth; in the 
calculations we assume that the medium remains transparent 
and discuss the conditions under which it becomes opaque in \S~6.1.)
Another important feature of GRBs is that
the radiation density strongly dominates over the rest-mass of the ambient 
medium, $F/c\gg \rho_0c^2$. The primary radiation dominates over the 
scattered radiation, $e^\pm$, and magnetic field in the front.

When the backreaction on the radiation pulse is negligible on 
time-scales $<R/c$, the propagating front is {\it quasi-steady}.
It can be formalized as follows. Let us define 
\begin{equation}
\vp=ct-R,
\end{equation}
where $t$ is the time passed since the beginning of the explosion.
Then we have $R=ct$ and $\vp=0$ at the leading boundary of the front,
and $\vp=\Delta$ at the trailing boundary.
Now let us change variables $(t,R)\rightarrow(t,\vp)$. That the front is 
quasi-steady means that the medium parameters are functions of $\vp\ll R$
and $t\approx R/c$ is a slowly changing parameter. 
The front gradually changes when its radius $R$ increases.
We aim to construct a model for the front structure, i.e. determine 
the medium density and velocity as functions of $\vp$.
We will show that the front evolves in a self-similar manner:
with increasing $R$, its structure is given by same unique functions of 
dimensionless coordinate $\xi=\vp/\lambda$.

The front is thin ($\Delta\ll R$) and its 
quasi-steady structure (formed on time-scales $\ll R/c$) can be described in 
plane-parallel geometry. The collimation of GRB radiation is very strong,
$\Delta\theta<0.01$, and to a first approximation it is perfectly collimated.
Asymmetry of the explosion does
not affect the calculations unless the ejecta are beamed within an angle
less than $\Delta\theta$. Possible inhomogeneity of the ambient medium is 
not considered in this paper [in contrast, Dermer \& B\"ottcher (2000) 
discussed the impact of the $\gamma$-rays on circumstellar clouds].

\subsection{Particle collectivization and the cold approximation}

It is reasonable to assume that the loaded $e^\pm$ share
immediately their net momentum with the medium.
Even though the Coulomb collisions are extremely inefficient,
there are two other mechanisms of momentum exchange: 

(1) The created $e^\pm$ form a stream interacting with the medium 
via beam instability. The instability time-scale is of order 
$\omega_{\rm pl}^{-1}$ where $\omega_{\rm pl}=(4\pi n_ee^2/m_e)^{1/2}$ 
is the plasma frequency and $n_e$ is the electron density. 

(2) In the presence of transverse magnetic field $B$ the pairs gyrate 
around the field lines frozen into the medium on the Larmor time, 
$\omega_B^{-1}=m_ec/Be$. The net momentum of $e^\pm$ is thus communicated 
to the medium.

The first mechanism should always work because its time-scale is shorter 
than the Compton cooling time of $e^\pm$, $t_{\rm C}=(3m_ec^2/8\sT F)$,
and also much shorter than the time of medium dynamics across the front.
The magnetic coupling may dominate if
$\omega_B>\omega_{\rm pl}$ which requires $B^2/4\pi>n_em_ec^2$. 
When the medium accelerates one should substitute the rest-frame 
magnetic field, density, and flux in these estimates.
The acceleration results in compression (Madau \& Thompson 2000),
the density and magnetic field are amplified, and the coupling becomes
even stronger.

The coupling passes the net momentum of injected $e^\pm$ to the medium and
maintains their distribution approximately isotropic in the medium rest frame.  
It does not ensure however that the $e^\pm$ also share their energy with other 
particles. One can distinguish between two possible situations:
(1) {\it Partial collectivization}: $e^\pm$ injected with a 
Lorentz factor $\gamma_e$ (measured in the medium rest frame) quickly get
isotropized but preserve $\gamma_e$; their subsequent cooling
is controlled by Compton losses. This is the case with magnetic coupling.
(2)~{\it Complete collectivization}: all particles share their energy 
instantaneously and maintain a Maxwellian distribution in the medium rest 
frame. It might be the case if collective modes provide sufficient coupling.

We will show that the majority of $e^\pm$ are loaded with moderately
relativistic energies, cool efficiently (even with partial collectivization), 
and remain at subrelativistic energies.
Particles created at $\vp$ are Compton cooled much faster than the medium 
moves to $\vp+a$ where next generation of hot pairs is created. Therefore
the bulk of pair-creating radiation at $\vp+a$ has been scattered by 
cooled particles.
To the first approximation, the medium can be considered
as a cold plasma with a bulk velocity $\beta$ found from momentum conservation. 
We hereafter use this ``cold'' approximation since it greatly simplifies
the calculations; its validity is checked in \S~4.


\section{Basic equations}

In this section we give the equations of a steady radiation front 
in the plane-parallel geometry (see \S~2.1).

\subsection{Scattering and pair creation}

Let $\mu$ be the cosine of the scattering angle. A primary collimated 
photon scattered through $\mu$ starts to move backward with 
respect to the $\Delta$-shell with velocity $\dd\vp/\dd t=c(1-\mu)$ and 
its $\vp$-coordinate grows. The scattering at $0<\vp^\prime<\vp$ determines
the intensity of scattered radiation at $\vp$,
\begin{equation}
\label{eq:scat}
  \Isc(\mu,\epsc)=\int_0^\vp \frac{\dd\vp^\prime}{1-\mu} \;
   F_\ep\, n\, \frac{(1-\beta)}{2\pi}
    \frac{\dd\sigma}{\dd\mu} \frac{\epsc}{\ep}\; e^{-\tgg}.
\end{equation}
Here $\dd\vp^\prime/(1-\mu)=c\dd t$ is the length element along the scattered 
ray, $\dd\sigma/\dd\mu$ is Compton cross-section (see Appendix),
$n$ is the electron/positron density, and $\beta$ is the medium velocity in
units of $c$. 
$F_\ep$, $n$, and $\beta$ are taken at the location of scattering, 
$\vp^\prime$. The photon energies before and after the scattering, $\ep$ and 
$\epsc$, are related by 
\begin{equation}
\label{eq:epsc}
  \epsc=\frac{\ep(1-\beta)}{1-\beta\mu+(1-\mu)\ep/\gamma},
\end{equation}
where $\gamma=(1-\beta^2)^{-1/2}$ is the Lorentz factor of the scattering
medium.

The scattered radiation that propagates from $\vp^\prime$ to $\vp$ is 
attenuated by $\gamma-\gamma$ absorption. This is accounted for by the 
exponential factor in equation~(\ref{eq:scat}) where $\tgg$ is the 
$\gamma-\gamma$ optical depth,
\begin{equation}
  \tgg=\int_{\vp^\prime}^\vp \kgg\dd\vp.
\end{equation}
The opacity $\kgg$ is dominated by the primary collimated
radiation (the scattered radiation has much smaller density, see \S~2.1).
A scattered photon $(\epsc,\mu)$ can interact with primary photons $\ep$
that are above the threshold 
\begin{equation}
\label{eq:thr}
  \epthr=\frac{2}{(1-\mu)\epsc}.
\end{equation}
The cross-section for interaction with $\ep\simgt\epthr$ is $\sgg\sim 0.1\sT$. 
The $\tgg$ can be viewed as the product of $(1-\mu)\sgg$ and the column 
density of primary photons above the threshold, 
$\sim s(F_{\epthr}/m_ec^3)$ where $s=c(t-t^\prime)=(\vp-\vp^\prime)/(1-\mu)$ 
is the path passed by the scattered photon. 

The exact expression for $\kgg(\mu,\epsc)$ is given in 
equation~(\ref{eq:opac}) of Appendix.
In numerical examples we will consider a homogeneous primary radiation pulse 
i.e. assume that the spectrum $F_\ep$ does not depend on $\vp$.
Then the $\gamma-\gamma$ opacity is homogeneous across the $\Delta$-shell
and $\tgg(\vp,\vp^\prime)=(\vp-\vp^\prime)\kgg$.

The pair creation rate at given $\vp$ is the rate of 
$\gamma-\gamma$ interaction between $\Isc(\mu,\epsc)$ and the primary beam
$F_\ep$,
\begin{eqnarray}
\nonumber
  \dot{n}_+(\vp) = 2\pi\int_{-1}^1\dd\mu\int\dd\epsc\; 
            \frac{\Isc(\mu,\epsc)}{\epsc m_ec^2}(1-\mu)\kgg\\
   =\int_0^\vp\dd\vp^\prime\int_{-1}^1\dd\mu\int
   \dd\ep\,\frac{\dd\epsc}{\dd\ep}\,
  \frac{F_\ep\kgg}{\ep m_ec^2} n(1-\beta)\frac{\dd\sigma}{\dd\mu}e^{-\tgg}.
\label{eq:ndot}
\end{eqnarray}
Here we made use of equation~(\ref{eq:scat}).

\subsection{Continuity equation}

Let $n_i$ and $n_e$ be the density of background ions and electrons,
and let $2n_+$ be the density of created $e^\pm$ pairs. The total electron 
density of the medium $n=n_e+2n_+$ and its velocity $v=\beta c$ satisfy the 
continuity equation. For a plane-parallel front the continuity equation reads
\begin{equation}
  \frac{\partial n}{\partial t}+
  \frac{\partial (nv)}{\partial R}=2\dot{n}_+-2\dot{n}_{\rm ann},
\end{equation}
where $\dot{n}_+$ and $\dot{n}_{\rm ann}$ are the local rates of 
pair creation and annihilation, respectively. 
The annihilation rate $\dot{n}_{\rm ann}=(3/8)(1-\beta^2)n_+^2\sT c$ is 
many orders of magnitude smaller than $\dnp$ 
and hereafter we neglect annihilation. 

Since both $n$ and $v$ are functions 
of $\vp=ct-R$ only, we have $\partial/\partial t=c\dd/\dd\vp$ and 
$\partial/\partial R=-\dd/\dd\vp$, and rewrite the continuity equation as
\begin{equation}
\label{eq:cont}
  c\frac{\dd }{\dd\vp}\left[n(1-\beta)\right]=2\dot{n}_+.
\end{equation}
The immediate consequence of this equation is that the magnitude 
\begin{equation}
\label{eq:nstar}
   n^*\equiv n(1-\beta) 
\end{equation}
would conserve in the absence of pair creation and hence the compression of 
accelerated medium is $(1-\beta)^{-1}$ (see also Madau \& Thompson 2000).
In particular, for the background electrons and ions we have
\begin{equation}
\label{eq:compr}
   n_e^*\equiv n_e(1-\beta)=n_0, \qquad
   n_i^*\equiv n_i(1-\beta)={n_i}_0.
\end{equation} 
Here $n_0$ and ${n_i}_0$ are the electron and ion densities prior to 
the interaction with the front.

The mass density of the medium is (we neglect the additional mass associated 
with the plasma internal energy: the cold approximation) 
\begin{equation}
\label{eq:ion}
  \rho=n_im_i+nm_e=\frac{\rho_0}{1-\beta}\left(1+\frac{n^*}{n_0}
   \frac{m_e}{\mu_em_p}\right), \quad \mu_e\equiv\frac{\rho_0}{n_0m_p}. 
\end{equation}
We neglected the small contribution 
($\sim m_e/m_p$) of the background electrons to $\rho_0$. 
The $m_i$ is the ion mass and $\mu_e$ is the medium mass (in units of $m_p$)
per electron: $\mu_e=1$ for hydrogen and $\mu_e=2$ for helium or heavier 
ions. The ratio
$n^*/n_0$ shows the number of $e^\pm$ loaded per one background electron. 

The cross-section for Compton scattering is 
inversely proportional to the squared mass of the scatter, so only $e^\pm$ are 
efficient scatters. The average mass per one scatter is
\begin{equation}
\label{eq:m}
  m_*=\frac{\rho}{n}.
\end{equation}
The initial $m_*=\mu_em_p$ can decrease to $m_e$ as a result of pair loading.

\subsection{Momentum conservation}

The law of momentum conservation reads (neglecting the pressure forces:
the cold approximation)
\begin{equation}
  \frac{\partial (v\gamma\rho)}{\partial t}+
  \frac{\partial (v^2\gamma\rho)}{\partial R}
  =\dot{P}_\pm+\dot{P}_{\rm sc},
\end{equation}
where 
$\dot{P}_\pm$ is the momentum deposited by pair creation per unit volume 
per unit time and $\dot{P}_{\rm sc}$ is the momentum deposited by
photon scattering off the medium. We rewrite this equation as
\begin{equation}
\label{eq:mom}
  c^2\frac{\dd }{\dd\vp}\left[\rho\beta\gamma(1-\beta)\right]
  =\dot{P}_\pm+\dot{P}_{\rm sc}.
\end{equation}
The scattering passes momentum from the beamed radiation to the medium 
with rate
\begin{equation}
\label{eq:Psc}
  \dot{P}_{\rm sc}=\left(1-\frac{\gamma^4}{\gsat^4}\right)\frac{n^*}{c}
                   \int\dd\ep\frac{F_\ep}{\ep}\int\dd\sigma(\ep-\mu\epsc).
\end{equation}
The factor $1-\gamma^4/\gsat^4$ accounts for a finite collimation angle of
the primary radiation (see eq.~\ref{eq:sat} of Appendix).
Assuming that the radiation is emitted by the ejecta with Lorentz
factor $\Gej$ at $R=\Rem$, we have $\gsat=\Gej (R/\Rem)$ at a radius $R$.

The momentum deposited by pair creation is given by
\begin{eqnarray}
\label{eq:Ppm}
 \dot{P}_\pm=\int_0^\vp\dd\vp^\prime\int_{-1}^1\dd\mu\int\dd\epsc\,
  \frac{F_\ep\kgg}{\ep m_ec^2} n^*\frac{\dd\sigma}{\dd\mu}e^{-\tgg} p_\pm.
\end{eqnarray}
Here $p_\pm(\mu,\epsc)$ is the average momentum of the $e^\pm$ pair 
created when a scattered photon $(\mu,\epsc)$ gets absorbed,
\begin{equation}
   \frac{p_\pm}{m_ec}=\mu\epsc+\chi\epthr.
\end{equation}
The numerical factor $\chi\sim 1$ is given in equation~(\ref{eq:chi}) of 
Appendix.

\subsection{Thermal balance}

The continuity and momentum equations allow one to compute the dynamics 
of the medium in the cold approximation. 
When we know the dynamics of the cold medium, we can 
evaluate its temperature from the thermal balance; it will allow us to check 
the consistency of the cold approximation. The thermal balance in the medium 
rest frame reads
\begin{equation}
\label{eq:law}
  \frac{\dd(u\tilde V)}{\dd\tilde t}=-p\frac{\dd\tilde V}{\dd\tilde t}
  +\ginj m_ec^2\frac{\dd(\tilde n\tilde V)}{\dd\tilde t}+(C^+-C^-)\tilde{V}.
\end{equation}
Here $u$ is internal energy density of the medium (including rest mass of 
$e^\pm$), $p$ is pressure, $\ginj(\vp) m_ec^2$ is the mean energy of 
injected $e^\pm$, $\tilde V$ is volume per barion, and
$\dd\tilde t=\dd t/\gamma$; all these magnitudes are measured in the rest
frame of the medium. 

The terms $C^\pm$ are the rates of Compton heating/cooling. 
Both depend on the particle energy distribution in the medium rest frame. 
Given the uncertainty of this distribution we replace it 
by $\delta$-function at a mean Lorentz factor $\gamma_e$ and estimate 
roughly
\begin{equation}
\label{eq:Compt}
  C^+-C^-=\frac{4}{3}\left(\beta_{\rm C}^2\gC^2-\beta_e^2\gamma_e^2\right)
          \tilde{n}\tilde{V}\sT\tilde{F}_T,
\end{equation}
where $\gamma_e=\gC$ corresponds to Compton equilibrium and
$\tilde{F}_{\rm T}$ is the flux of (primary) radiation that scatters in 
Thomson regime; this flux is measured in the medium rest frame and it is 
approximately
\begin{equation}
  \tilde{F}_{\rm T}=\FT(\ep<\epKN)\frac{1-\beta}{1+\beta}, \qquad 
  \epKN\sim \frac{\gamma(1+\beta)}{\gamma_e}.
\end{equation}
Here $\epKN$ is the typical energy above which the scattering occurs in the
Klein-Nishina regime. $\FT(\ep<\epKN)$ is the primary flux 
with $\ep<\epKN$, measured in the lab frame.

The ions carry a small fraction of the thermal energy 
(even if they manage to share the energy with $e^\pm$, their density 
$n_i\ll n$ as soon as pair creation begins) and hence 
$u\approx\gamma_em_ec^2\tilde{n}$. 
Note that $\tilde{n}\tilde{V}=n^*$ and $\tilde{V}=\gamma(1-\beta)V_0$
where $V_0$ is volume per barion prior to the interaction with the front. 
Substituting these relations into equation~(\ref{eq:law}) and taking into 
account that $\dd\vp/\dd\tilde{t}=c\gamma(1-\beta)$ we get after simple 
algebra
\begin{eqnarray}
\nonumber
   \frac{\dd\gamma_e}{\dd\vp}=-\frac{(p/\tilde{n}m_ec^2)}{\gamma(1-\beta)}
    \frac{\dd}{\dd\vp}\left[\gamma(1-\beta)\right]
   +\frac{(\ginj-\gamma_e)}{n^*}\frac{\dd n^*}{\dd\vp} \\
   +\frac{4}{3}\frac{(\gC^2-\gamma_e^2)}{\gamma(1+\beta)}
    \frac{\sT\FT}{m_ec^3}. \hspace*{1cm}
\label{eq:thermal}
\end{eqnarray}
We estimate the pressure interpolating between nonrelativistic 
$p/\tilde{n}m_ec^2=(2/3)(\gamma_e-1)$ and relativistic 
$p/\tilde{n}m_ec^2=\gamma_e/3$ limits, 
\begin{equation}
\label{eq:pressure}
  \frac{p}{\tilde{n}m_ec^2}\approx\frac{\gamma_e^2-1}{3\ge}
   \approx\frac{kT}{m_ec^2}\equiv\Theta.
\end{equation}
Here we introduced a temperature $T$. For a Maxwellian plasma
$T$ is related to pressure by $p=\tilde{n}kT$ in both non-relativistic and 
relativistic cases. For a non-Maxwellian distribution, $T$ is an effective
temperature defined by $p=\tilde{n}kT$.

Once we know $\gamma(\vp)$ and $n^*(\vp)$ from the dynamic ``cold'' solution
we 
can find $\FT$ and $\ginj(\vp)$ (see Appendix). Then we can solve numerically 
equation~(\ref{eq:thermal}) and find $\gamma_e(\vp)$ and $\Theta(\vp)$.


\section{Numerical results}

In this section we construct a numerical model of the front. 
Here we assume that the ambient medium is hydrogen ($\mu_e=1$, 
see eq.~\ref{eq:ion}); the extension to $1<\mu_e<2$ is simple (\S~5).

We integrate the ordinary differential equations~(\ref{eq:cont}) and
(\ref{eq:mom}) with the boundary conditions $\beta=0$ and $n=n_0$ at $\vp=0$.
At each step $\dd\vp$ we know the radiation scattered at previous steps
(smaller $\vp$) and find the local 
pair creation rate from equation~(\ref{eq:ndot}) and the rate of momentum
injection from equations~(\ref{eq:Psc}) and (\ref{eq:Ppm}).
After getting the dynamic solution $n(\vp)$ and $\gamma(\vp)$ we integrate
the thermal balance equation~(\ref{eq:thermal}) with the boundary condition
$\gamma_e(0)=1$ and find $\gamma_e(\vp)$. 

The input of the calculations is the GRB spectrum $F_\ep(\vp)$ and the 
output is the front structure $n(\vp)/n_0$, $\beta(\vp)$, and $\ge(\vp)$.
For numerical illustration we take a radiation pulse with constant spectrum
\begin{equation}
\label{eq:spectr}
   F_\ep=
  \left\{\begin{array}{ll}
  F_1\ep^{-\alpha_1}, & \ep<1,\\ 
  F_1\ep^{-\alpha_2}, & 1<\ep<\epbr,\\
  0, & \ep>\epbr.\\
  \end{array}\right.
\end{equation}
Such a spectral shape is observed in GRBs with $\alpha_1\sim \pm 0.5$
and $\alpha_2\sim 1.5\pm 0.5$ (Preece et al. 2000). 
In numerical examples we fix $\alpha_1=0$ and assume $\alh>1$. 
Then the problem has a well defined solution in the limit 
$\epbr\rightarrow \infty$. The finiteness of $\epbr\gg 1$ causes a break in 
pair loading at $\vp=\vpbr$ (see \S\S~5.3 and 6.2). In the examples below we 
take $\epbr=10^2$. 

The solution is a function of the dimensionless 
coordinate\footnote{ 
The $\xi$-coordinate has the meaning of dimensionless fluence of the burst, 
$\xi=(\sT/m_ec^3)F\vp$. The computed $n(\xi)$ are $\gamma(\xi)$ are also the
exact solution for bursts with arbitrary light curves $F_1(\vp)$ (but with a 
fixed spectral shape) once $\xi$ is defined as 
$\xi=(\sT/m_ec^3)\int_0^\vp F\dd\vpp$.
}
$\xi=\vp/\lambda$
where $\lambda$ is given by equation~(\ref{eq:lam}) with the total flux 
\begin{equation}
\label{eq:F}
  F=\frac{(\alpha_2-\alpha_1)F_1}{(1-\alpha_1)(\alpha_2-1)}.
\end{equation}

\subsection{Dynamical structure of the front}

Figure~1 shows the computed front structure in the case of 
$\alpha_2=1.5$ and $\gsat=10^3$ (the solution does not depend on $\gsat$ 
until $\gamma$ approaches $\gsat$, cf. eq.~\ref{eq:Psc}).
Near the leading boundary of the front,
the medium has $\gamma\approx 1$ and pair loading proceeds 
exponentially on scale of $\xiload\approx 30$. At $\xiacc\approx 10^2$ 
the medium accelerates. 

The acceleration length can be understood in simple terms.
When a portion $\dd\vp$ of the radiation pulse overtakes an electron with
$\gamma\approx 1$, it passes momentum $\dd p\approx 0.2(F\sT/c^2)\dd\vp$
(here 0.2 is a Klein-Nishina correction).
Hence $\dd p/\dd\xi\approx 0.2m_ec$ and the medium acceleration length 
is $\xiacc\approx 5m_*/m_e$ where 
$m_*=m_p(n_0/n^*)$ is mass per scatter (see eqs.~\ref{eq:m} and 
\ref{eq:ion}). This yields an estimate
\begin{equation}
\xiacc\approx 5(m_p/m_e)\exp(-\xiacc/30),
\end{equation}
i.e. $\xiacc\approx 10^2$. The estimate neglects the additional acceleration 
due to $\Ppm$ which is approximately equal to $\Psc$ at $\xi\sim \xiacc$
(see Fig.~2). More exact formulae are derived in \S~5.

The pair loading continues in the accelerated zone $\xi>\xiacc$ and $m_*$ 
further decreases. Therefore
the medium accelerates very fast, $\gamma\approx (\xi/\xiacc)^3$,
until $n^*/n_0$ reaches $m_p/m_e$ and $m_*$ saturates at $m_e$;
afterward $\gamma\propto \xi^{3/2}$. 

The accelerating medium scatters radiation through smaller angles, 
$\mu\approx\beta\rightarrow 1$, and pair loading slows down: 
$\dd^2 n^*/\dd \xi^2$ becomes negative at $\xi\approx\xiacc$. 
The decrease in pair production is caused by the growth of the threshold for 
$\gamma-\gamma$ interaction, $\epthr\propto(1-\mu)^{-1}$ (eq.~\ref{eq:thr}).
The scattering by medium with a relativistic $\gamma$ produces photons
with $1-\mu\approx 1/2\gamma^2$, and hence $\epthr\propto\gamma^2\propto 
(\xi/\xiacc)^6$. The $\gamma-\gamma$ opacity seen by the beamed scattered 
photons, $\kgg\propto \epthr^{-\alh}$, becomes very low and they travel 
almost freely across the front with a free-path 
$\lgg=\kgg^{-1}\propto(\xi/\xiacc)^{6\alh}$.
For instance photons with $\ep=1$ scattered by medium with $\gamma=2$ 
(at $\xi=1.3\xiacc$) are absorbed only at $\xi\sim 10^4$, i.e. absorption 
strongly lags behind scattering. It implies that photons scattered at
$\xi\simgt \xiacc$ control the pair loading in the whole relativistic zone
of the front. The pair loading thus decouples from the medium dynamics at 
$\xi\gg\xiacc$. As explained in \S~5, the bulk of radiation scattered at 
$\xi>\xisc\approx 1.4(\epbr/100)^{1/6}$ (where $\gamma>\gsc=\sqrt{\epbr}/4$) 
is not absorbed at all and escapes the front.

At $\xi_\pm\approx 10^3$ the $e^\pm$ density exceeds the density of
the ambient electrons by the factor $m_p/m_e$ and $m_*$ saturates at $m_e$. 
Then equations~(\ref{eq:mom}) and (\ref{eq:Psc}) yield
\begin{equation}
\label{eq:accel}
  \frac{\dd\gamma}{\dd\xi}=\frac{1}{2}\left(1-\frac{\gamma^4}{\gsat^4}\right)
  -\frac{\gamma}{n^*}\frac{\dd n^*}{\dd\xi}.
\end{equation}
Here we took $\beta\approx 1$ and calculated $\Psc$ with Thomson 
cross-section (medium with $\gamma\gg 1$ scatters the GRB radiation
in Thomson regime). We neglected $\Ppm$ compared to $\Psc$,
which is a good approximation at $\gamma\gg 1$ (see Fig.~2).
In the absence of pair loading ($\dd\ln n^*/\dd\ln\xi\ll 1$), $\gamma$ would 
tend to the asymptotics $\gamma=\xi/2$. However before $\gamma$ can reach 
any asymptotics it saturates. The saturation happens at 
$\xi=\xisat\simlt 10^4$.

\centerline{ \epsfxsize=9.0cm {\epsfbox{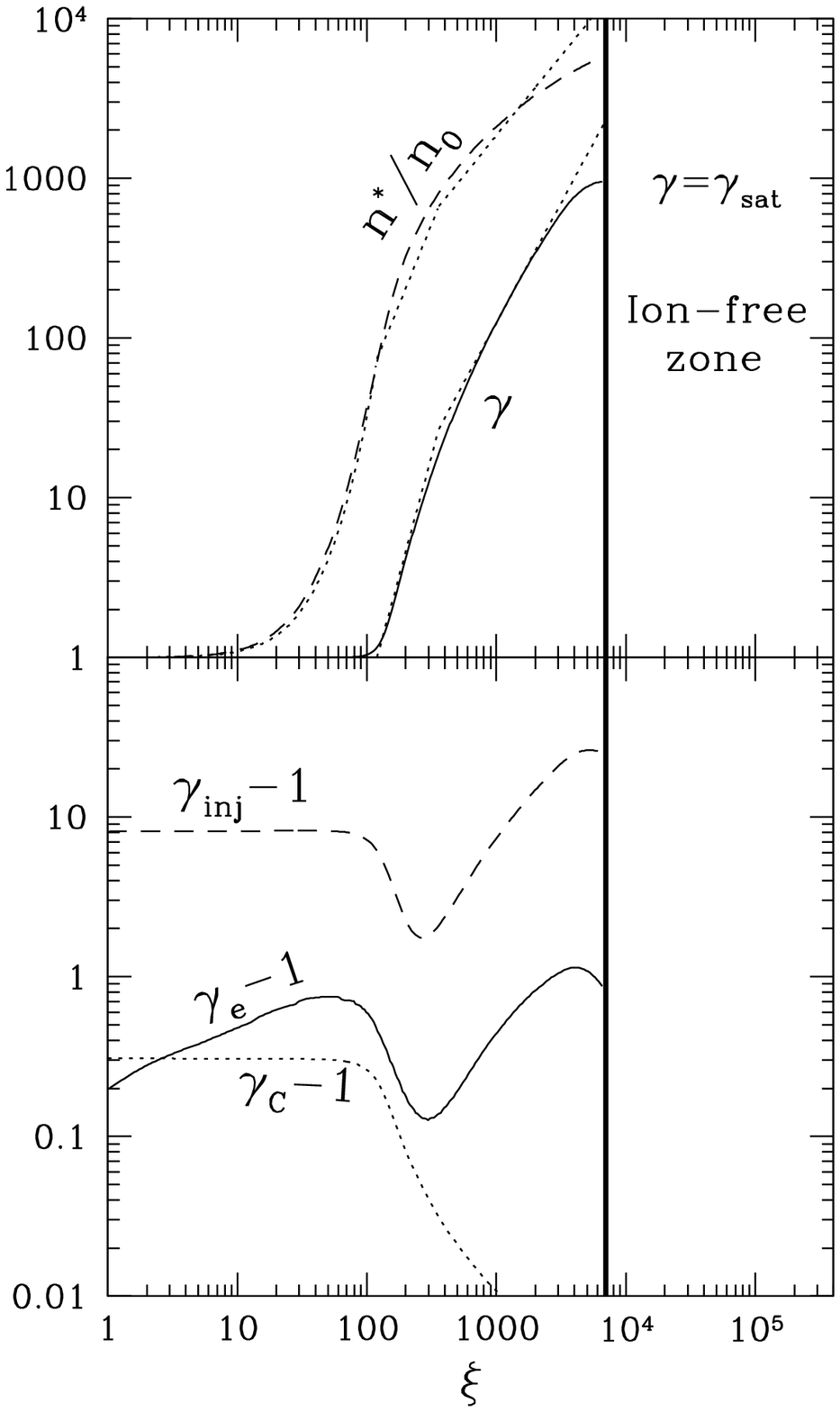}} }
\figcaption{ 
Structure of the radiation front for $\als=0$, $\alh=1.5$, $\epbr=10^2$.
{\it Top}: Dynamic structure. {\it Dashed} and {\it solid curves} show 
$n^*(\xi)/n_0$ and $\gamma(\xi)$ where $\xi=\vp/\lambda$.
{\it Solid vertical line} shows the boundary of the ion-free zone, $\xi_c$,
where $\gamma$ reaches $\gsat$. {\it Dotted curves} show the analytical model
of \S~5 (see eqs.~\ref{eq:stat},\ref{eq:ac},\ref{eq:appr1},\ref{eq:appr2}). 
{\it Bottom}: Thermal structure. {\it Solid curve} shows the 
mean kinetic energy of particles in the medium rest frame.
The other two curves display $\ginj-1$ and $\gamma_{\rm C}-1$
(cf. the text).
}
\bigskip

Our steady dynamic problem becomes inconsistent when $\gamma$ saturates.
The assumption that the front has the speed of light and the medium passes 
through it with $\dd\vp/\dd t=1-\beta$ becomes wrong. Instead the 
medium gets stuck in the front: it has reached the velocity $\beta_{\rm sat}$
such that the net flux of GRB radiation vanishes in the medium rest frame. 
The $\beta_{\rm sat}$ is determined by the angular spread of the 
radiation and represents the effective velocity of the radiation pulse.
Saturation implies that the ambient (ion) medium is trapped in the pulse
and cannot penetrate the zone $\xi>\xisat$ --- this zone is ion-free. 
(More exactly, the ions cannot penetrate $\xi>\max\{\xi_c,\ximix\}$ where
$\ximix\sim \gsat^{-2}R/\lambda$, see \S~7.1.)
The trapped ions accumulate and surf the pulse.

\centerline{ \epsfxsize=9.5cm {\epsfbox{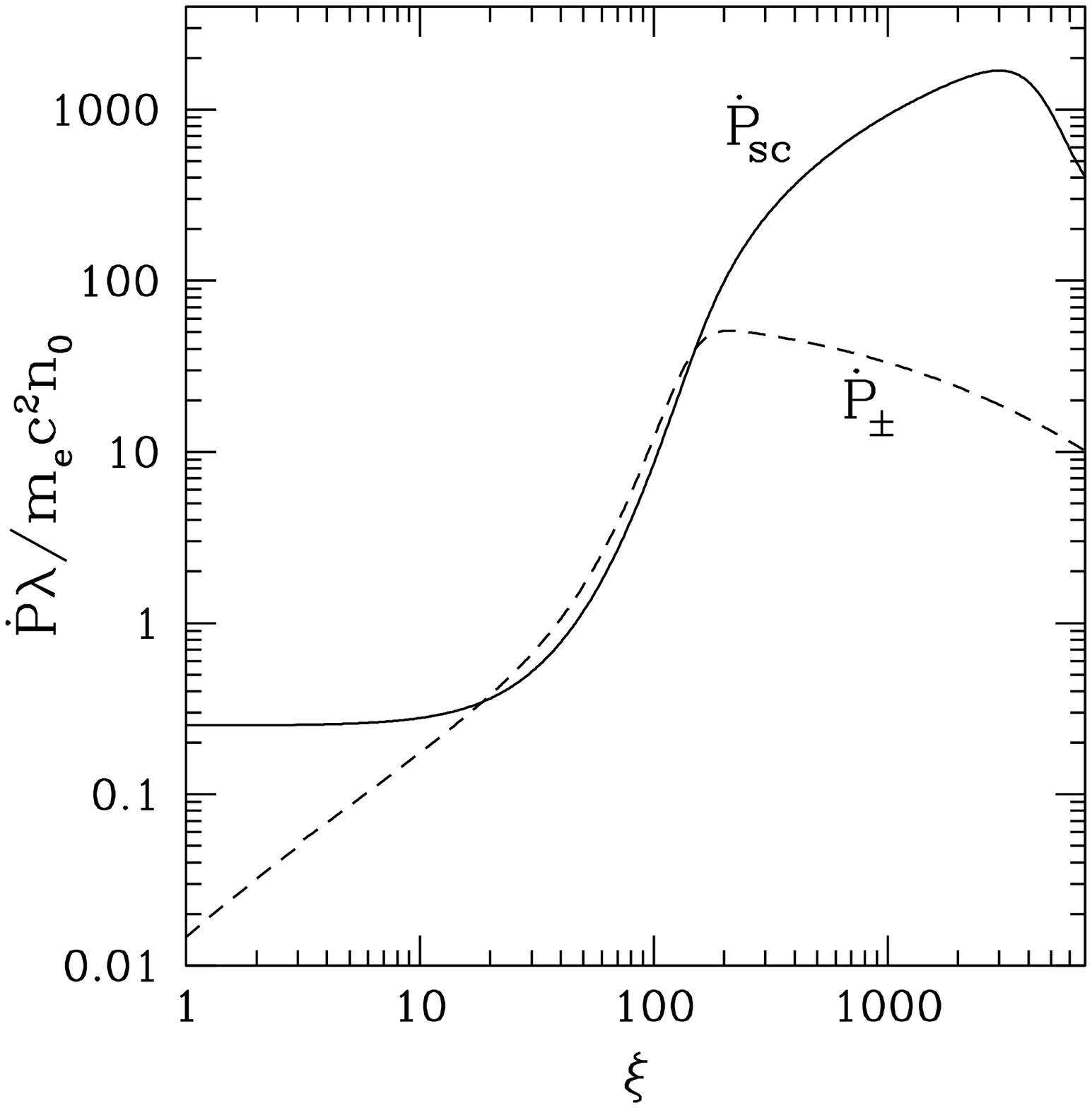}} }
\figcaption{ Momentum deposition rate in the front shown in Figure~1. 
{\it Solid} and {\it dashed curves} display $\Psc$ and $\Ppm$
(see eqs.~\ref{eq:mom}-\ref{eq:Ppm}).}
\bigskip

Radiation scattered by the medium in the process of its acceleration partially
propagates to the ion-free zone of the front and produces $e^\pm$ there 
owing to $\gamma-\gamma$ reaction with the primary radiation.
A steady $\dnp(\vp)$ is established throughout the whole front 
on a relatively short time-scale $\sim \vp/(1-\mu)c$ where 
$1-\mu\simlt(1/2\gsc^2)\sim 0.1$ represents the typical collimation angle 
of the scattered radiation that produces pairs.
The pairs created in the ion-free zone acquire the saturated Lorentz factor,
stay almost static in the $\vp-$coordinate, and accumulate.

The constructed plane-parallel model can be applied to the
expanding spherical front as long as the time-scales involved do not exceed
$R/c$. In particular, the time $\tacc(\gamma)$ of medium acceleration 
to a given $\gamma(\vp)$ should be smaller than $R/c$. To estimate $\tacc$
assume the most favorable conditions for acceleration: $m_*=m_e$
and $\dd\ln n^*/\dd\ln\xi\ll 1$; then equation~(\ref{eq:accel}) reads 
$\dd\gamma/\dd t=(c/\lambda)(1-\beta)/(1+\beta)\approx
(c/4\gamma^2\lambda)$ and gives $\tacc\approx \gamma^3\lambda/c$. 
From $\tacc<R/c$ we find $\gamma<\gmax$ where
\begin{equation}
\label{eq:gmax}
  \gmax\approx \left(\frac{R}{\lambda}\right)^{1/3}
   = 6\times 10^2R_{15}^{-1/3}L_{53}^{1/3}.
\end{equation}
Both $\gsat$ and $\gmax$ evolve as the spherical front expands. 
If $\gmax<\gsat$ the boundary of the ion-free zone $\xisat$ is determined 
by $\gamma=\gmax$ rather than $\gamma=\gsat$. 
We discuss the front evolution in more detail in \S~7 and
derive $\xi_c(R)$ there.
Here it is worth to emphasize that the front structure shown in Figure~1
at $\xi<\xi_c$ does not change with radius, i.e. the front is self-similar.
Since $\lambda\propto R^2$ the front is just ``stretched'' in $\vp$-coordinate 
as $R$ grows. Note also that 
the front trailing boundary $\xiD=\Delta/\lambda$ in Figure~1 
moves to the left with increasing $R$ and becomes smaller than $\xi_c$ at
some radius $R_c$ found in \S~7. At $R>R_c$ there is no ion-free zone and 
the whole front is described by the self-similar solution.

\subsection{Thermal structure of the front}

In the bottom panel of Figure~1 one can see two peaks of $\gamma_e$ at 
$\xi\approx 70$ and $\xi\approx 4\times 10^3$. They correspond to the beginning 
and the end of the medium acceleration. This unusual temperature profile can 
be understood from equation (\ref{eq:thermal}) which we rewrite as
\begin{equation}
\label{eq:cool}
  \frac{\dd\gamma_e}{\dd\xi}=\frac{\Theta}{\xiacc}
                             +\frac{\ginj-\ge}{\xiload}-\frac{\ge}{\xiC}.
\end{equation}
Here the effective temperature $\Theta$ is related to the average Lorentz 
factor $\ge$ via equation~(\ref{eq:pressure}), 
$\xiacc=(-\dd\ln[\gamma(1+\beta)]/\dd\xi)^{-1}$ is the acceleration length, 
$\xiload=(\dd\ln n^*/\dd\xi)^{-1}$ is the pair loading length, and 
\begin{equation}
  \xiC=\frac{3}{4}\frac{F}{\FT}\frac{\gamma(1+\beta)\ge}{\ge^2-\gC^2}=
  \left\{\begin{array}{ll}
  \frac{3}{4}\frac{F}{\FT}\frac{\gamma(1+\beta)}{\ge}, & \ge\gg 1,\\ 
  \frac{3}{4}\frac{F}{\FT}\frac{\gamma(1+\beta)}{\beta_e^2-\beta_{\rm C}^2}, 
     & \beta_e\ll 1,\\ 
  \end{array}\right.
\end{equation}
is the length of Compton cooling. 

The initial temperature of the medium is low and it gradually rises
at small $\xi$ owing to injection of pairs with $\ginj\sim 10$.
Already at $\xi\approx 3$ the temperature 
exceeds the Compton equilibrium value $\ThC$. Thereafter $\Theta>\ThC$ and 
Compton scattering cools the medium rather than heats. 
At $\xi\approx\xiload\sim 30$ the pair density exceeds that of 
background electrons and begins to exponentiate. One could then expect 
a high heating rate, however Compton cooling is very efficient and keeps 
the temperature below $m_ec^2$. 
The length of Compton cooling is $\xiC\approx (F/\FT)\epKN\sim 1$.
It is much shorter than $\xiload$ and therefore the cooling competes 
successfully with the heating. This competition is described by equation
\begin{equation}
\label{eq:cool1}
  \frac{\dd\gamma_e}{\dd\xi}\approx\frac{\ginj}{\xiload}-\frac{\ge}{\xiC}.
\end{equation}
Here we neglected the first (adiabatic) term on the right-hand side of 
equation~(\ref{eq:cool}) since it is much smaller than the other two terms.
Equation~({\ref{eq:cool1}}) shows that $\ge-1$ saturates at 
$\sim \xiC\ginj/\xiload\simlt 1$. This is the first maximum of the temperature
profile.

At $\xi\approx\xiacc\approx 10^2$ the medium begins to accelerate
and then the relative velocity between the injected 
$e^\pm$ stream and the medium decreases. Correspondingly, $\ginj$,
the heating rate, and the medium temperature fall down. 

When the medium Lorentz factor reaches $\gamma\sim 10$, 
the relative velocity between the injected $e^\pm$ stream and the medium
vanishes and changes sign. Here $\ginj$ reaches a minimum.
Afterward the $e^\pm$ loading tries to {\it decelerate} the medium 
(see also \S~5.4). The acceleration by scattering, however, dominates
and the medium continues to accelerate. 
Now $\ginj$ rises again (the relative velocity between the injected $e^\pm$
and the medium again increases) and the heating rate and the temperature 
grow.

The cold approximation is especially good near the minimum of $\ge$ at about
$2\xiacc$. The temperature is also quite low at $\xiacc<\xi<2\xiacc$
where the main scattering occurs (that controls pair loading 
in the whole accelerated zone of the front).

The energy distribution of $e^\pm$ around the average $\gamma_e$ depends
on details of their thermalization. In the case of partial collectivization
(see \S~2.2) the distribution has a tail extending from $\gamma_e$ to $\ginj$
whose slope is controlled by Compton cooling.
The length-scale for Compton cooling of injected $e^\pm$ is 
$\xiC(\ginj)\sim (F/\FT)\epKN\sim 1$ (it does not depend on $\ginj$ or 
$\gamma$ because $\FT/F\sim\epKN$ in the case of $\alpha_1=0$). 
Hence the density of pairs with $\gamma\sim\ginj$ is 
\begin{equation}
  n^*_{\rm inj}\sim \frac{\dd n^*}{\dd\xi}=\frac{n^*}{\xiload},
\end{equation}
i.e. the number density of the high-energy particles is about 30
times smaller as compared to cooled particles. 
Also their energy density is smaller than that of the cooled component. 

We conclude that the cold approximation is reasonably good. Note however
that we focus on relatively soft spectra $\alh>1$ (in contrast, TM 
took $\alh=1$ as a basic case). The case of hard spectra $\alh\simlt 1$ 
is more complicated because the maximum of $\gamma_e$ at $\xi\simlt \xiacc$
becomes essentially relativistic and then numerical simulations relaxing 
the cold approximation will be needed. A relativistically hot plasma scatters 
preferentially backward (smaller $\mu$);
note also that $\Ppm$ then strongly dominates over $\Psc$ at $\xi\sim\xiacc$. 
We expect however that the front structure will not change 
qualitatively for hard spectra, though the values of $\xiload$ and $\xiacc$
may change.


\section{Analytical model}

The medium dynamics in the radiation front can be understood 
with a simplified model that we formulate below. In particular, we derive 
the characteristic lengths $\xiload$ and $\xiacc$, get an analytical 
solution for the front in the non-relativistic ($\beta<0.5$) zone, 
and evaluate the pair loading rate in the accelerated zone.

\subsection{Formulation}

Let us replace the scattering cross-section by 
\begin{equation}
\label{eq:toy1}
   \frac{\dd\sigma}{\dd\mu}=\sT\delta(\mu-\beta)H(\epKN-\ep),
\end{equation}
where $\delta$ is the Dirac function and $H$ is the Heaviside step function.
Here we have made two approximations:

\begin{enumerate}

\item
Assume that radiation scatters with Thomson cross-section if $\ep<\epKN$
and does not scatter at all if $\ep>\epKN$, where $\epKN\simlt 1$ is the 
energy above which the Klein-Nishina corrections reduce the 
scattering and subsequent pair creation. We derive in Appendix the 
effective 
\begin{equation}
\label{eq:epKN}
   \epKN\approx 0.4\gamma(1+\beta)
\end{equation}
for calculations of $\dnp$ and $\Ppm$, and 
\begin{equation}
\label{eq:epKN_acc}
   \epKN^{\rm acc}\approx 0.7\gamma(1+\beta)
\end{equation}
for calculations of $\Psc$.

\item
Replace the broad distribution of the scattering angles by its average, 
$\mu=\beta$, i.e. assume that the collimated radiation scatters through 
90$^{\rm o}$ ($\tilde\mu=0$) in the medium rest frame. Then we also have
\begin{equation}
\label{eq:toy2}
  \epsc=\frac{\ep}{1+\beta}.
\end{equation}

\end{enumerate}

The scattered photons can interact with primary photons of energy 
$\ep>\epthr$ where the threshold is given by equation~(\ref{eq:thr}).
In our simplified model equation~(\ref{eq:thr}) reads
\begin{equation}
\label{eq:thr1}
  \epthr=\frac{2(1+\beta)}{\ep(1-\beta)}.
\end{equation}
We have $\epthr>1$ for any $\ep<\epKN$, i.e. the scattered 
radiation interacts with the high-energy part of the spectrum, 
$F_\ep=F_1\ep^{-\alh}$, $\ep>1$.
The $\gamma-\gamma$ opacity of the power-law radiation seen by the scattered 
photon is (eq.~\ref{eq:opac_pw} in Appendix)
\begin{equation}
\label{eq:toy3}
  \kgg=\frac{\ph(\alh)}{\l1}\left(\frac{\epthr}{2}\right)^{-\alh}\,
       H(\epthr-\epbr), \quad \lambda_1=\frac{m_ec^3}{F_1\sT}.
\end{equation}
The numerical factor $\ph(\alpha)$ can be approximated with high accuracy as 
(Svensson 1987)
\begin{equation}
\label{eq:phi}
  \ph(\alpha)\approx \frac{7}{12}2^{-\alpha}(1+\alpha)^{-5/3}.
\end{equation}
Hereafter we use notation $\phi\equiv\ph(\alh)=$0.045 and 0.023 for
$\alh=$1.5 and 2 respectively.
Equation~(\ref{eq:toy3}) is exact for a power-law spectrum and
inaccuracies appear only when $\epthr$ approaches the spectral break $\epbr$.
Given the opacity, we also know the free path of the scattered photons
$\lgg=\kgg^{-1}$,
\begin{equation}
\label{eq:lgg}
   \lgg=\frac{\l1}{\phi}\left(\frac{\epthr}{2}\right)^{\alh}.
\end{equation}

Finally, let us replace the exponential attenuation of the scattered radiation
in equation~(\ref{eq:scat}) by the step function $H(1-\tgg)$. Then 
equations~(\ref{eq:ndot}), (\ref{eq:Ppm}), and (\ref{eq:Psc}) read 
\begin{eqnarray}
\nonumber
  \dot{n}_+=\frac{\phi c}{\lambda_1^2}\int_0^\vp\dd\vp^\prime\,
    \frac{n^*(\vpp)}{1+\bp} 
     \left(\frac{1-\bp}{1+\bp}\right)^{\alh}
      \int_0^{\epKN} \dd\ep\,f_\ep\\
   \times \ep^{\alh-1}H(\epmax-\ep) H(\ep-\epmin),
\label{eq:ndot1}
\end{eqnarray}
\begin{eqnarray}
\nonumber
  \Ppm=\frac{\phi c}{\lambda_1^2}\int_0^\vp \dd\vp^\prime\,
    \frac{n^*(\vpp)}{1+\bp} 
     \left(\frac{1-\bp}{1+\bp}\right)^{\alh}
      \int_0^{\epKN} \dd\ep\,f_\ep \\
   \times \ep^{\alh-1}p_\pm\, H(\epmax-\ep) H(\ep-\epmin),
\label{eq:Ppm1}
\end{eqnarray}
\begin{equation}
\label{eq:Psc1}
  \Psc=\left(1-\frac{\gamma^4}{\gsat^4}\right)
            \frac{n^*m_ec^2}{(1+\beta)\l1}\int_0^{\epKN^{\rm acc}}f_\ep\dd\ep.
\end{equation}
Here $f_\ep=F/F_1=\ep^{-\als}$~ if~ $\ep<1$ and $f_\ep=\ep^{-\alh}$~ if~ 
$\ep>1$,
$\epmin$ is found from the condition $\epthr<\epbr$,
\begin{equation}
\label{eq:epmin}
  \epmin(\vp^\prime)=\frac{2}{\epbr}\left(\frac{1+\bp}{1-\bp}\right),
\end{equation}
and $\epmax$ is found from the condition $\tgg=(\vp-\vp^\prime)\kgg<1$,
\begin{equation}
\label{eq:epmax}
  \epmax(\vp^\prime)=
  \left[\frac{\lambda_1}{(\vp-\vp^\prime)\phi}\right]^{1/\alpha_2}
         \left(\frac{1+\bp}{1-\bp}\right).
\end{equation}
When $\epmax>\epbr$ one should replace the upper limit by $\epbr$.
This refinement is however not important since $\epbr$ is anyway far from
the scattered peak $\ep\sim 1$ that dominates pair loading as shown below.

The formula for $p_\pm$ (eq.~\ref{eq:gpm} of Appendix) in our simplified 
model reads
\begin{equation}
\label{eq:ppm1}
  \frac{p_\pm}{m_ec}=\frac{\ep\bp}{1+\bp}
   +\frac{(1+\bp)}{\ep(1-\bp)}\frac{\ph(\alh-1)}{\ph(\alh)}.
\end{equation}

The $\ep$-integrals in equations~(\ref{eq:ndot1}), (\ref{eq:Ppm1}), and 
(\ref{eq:Psc1}) depend on the relative positions of $\epmin$, $\epmax$,
$\epKN$, and unity. We now consider two different zones of the front
starting from small $\vp$.

\subsection{Non-relativistic zone ($\beta\ll 1$)}

In the non-relativistic zone we have $\epmin\ll\epKN<1<\epmax$ and 
equations~(\ref{eq:cont}) and (\ref{eq:ndot1}) give
\begin{eqnarray}
\nonumber
  \frac{\dd n}{\dd\vp}=\frac{2\dot{n}_+}{c}
   =\frac{2\phi}{\l1^2}\int_0^\vp \dd\vp^\prime\,n(\vpp)
      \int_{\epmin}^{\epKN}\ep^{\alh-1}f_\ep\dd\ep \\
   =\frac{2\phi \epKN^{\alh-\als}}{\l1^2(\alh-\als)}
     \int_0^\vp n(\vpp)\,\dd\vpp.
\label{eq:cont_st}
\end{eqnarray}
Here we neglected $\epmin$ compared to $\epKN$. 
The exact solution of equation~(\ref{eq:cont_st})
is the sum of growing and decaying exponentials,
\begin{equation}
\label{eq:stat}
   n=\frac{n_0}{2}\left(e^{\vp/\vpload}+e^{-\vp/\vpload}\right), \quad
   \frac{\vpload}{\l1}=\left(\frac{\alh-\als}{2\phi\epKN^{\alh-\als}}
                       \right)^{1/2}.
\end{equation}
Substituting $\als=0$ and $\epKN=0.4$ we get $\xiload=a/\lambda\approx
24$ and 33 for $\alh=1.5$ and 2 respectively
(here we used $\l1/\lambda=F/F_1$ given by eq.~\ref{eq:F}).
The analytical solution is in perfect agreement with the numerical results, 
see Figures~1 and 3.

The loading length admits an easy interpretation.
As seen from equation~(\ref{eq:cont_st}), scattered photons with 
$\ep\sim\epKN$ make the dominant contribution to $\dnp$.
Equations~(\ref{eq:lgg}) and (\ref{eq:thr1}) give the 
free-path of these photons,
\begin{equation}
\label{eq:lgg_st}
   \lgg\approx \frac{\l1}{\phi}\epKN^{-\alh}.
\end{equation}
One can see that $a\approx \sqrt{\lgg\l1}\approx \sqrt{\lgg\lambda}$.
Note that 
$\lgg/\lambda\sim 200$ and 500 for $\alh=1.5$ and 2 respectively,
i.e. the scattered radiation is weakly absorbed in the non-relativistic zone
(this is a consequence of $\epKN<\epmax$). 
When an ambient electron has passed a distance $\vp$ through the front, 
it has scattered $\sim \vp/\lambda$ photons and a fraction
$\sim \vp/\lgg$ of these photons have been absorbed. Hence one pair
is injected per one ambient electron when $(\vp/\lambda)\times(\vp/\lgg)=1$
which gives the above formula for the loading length $a$.

We now evaluate the medium acceleration at $\beta\ll 1$. First let us 
calculate the momentum loaded by $e^\pm$ pairs.
Substituting $p_\pm$ from equation~(\ref{eq:ppm1})
into equation~(\ref{eq:Ppm1}) we get
\begin{eqnarray}
\nonumber
  \Ppm=\frac{\phi m_ec^2}{\l1^2}\int_0^\vp \dd\vp^\prime\, 
   n(\vpp) \int_{\epmin}^{\epKN} \dd\ep\,\ep^{\alh-2}f_\ep 
   \frac{\ph(\alh-1)}{\ph(\alh)}\\
  = \frac{\ph(\alh-1) m_ec^2 \epKN^{\alh-\als-1}}{2\l1^2(\alh-\als-1)}
   \,n_0a\left(e^{\vp/a}-e^{-\vp/a}\right). \hspace*{0.5cm}
\label{eq:Ppm_st}
\end{eqnarray}
This is a perfect approximation if $\epbr\rightarrow\infty$. The 
$\ep$-integral in (\ref{eq:Ppm_st})
peaks at the upper limit as $\epKN^{\alh-\als-1}$.
One can take $\ep_*\sim\epKN/2$ as a typical $\ep$ of scattered photons,
then one gets the typical energy of absorbed primary photons 
$\epabs\approx\chi\epthr\approx(1+\alh^{-1})^{5/3}(2/\ep_*)\approx 20$,
assumed to be well below $\epbr$. 
Note that $\alh-\als-1\rightarrow 0$ when $\alh\rightarrow 1$ and $\als=0$,
i.e. for hard spectra the $\epsilon$-integral in (\ref{eq:Ppm_st}) does 
not have a pronounced peak and low-energy photons $\ep\ll\epKN$ contribute 
a lot to $\Ppm$. Such scattered photons interact with very energetic primary 
photons $\epabs\approx 10\epKN/\ep$ and then the finiteness of $\epbr$ is 
important: $\epabs>\epbr$ is excluded which leads to a reduction of $\Ppm$.
E.g. in the case of $\alh=1.5$ and $\epbr=10^2$ 
the actual $\Ppm$ is suppressed by a factor of 2 
compared to equation~(\ref{eq:Ppm_st}).

Equation~(\ref{eq:Psc1}) gives a perfect approximation to the momentum 
deposited by scattering. Where $\beta\ll 1$ it yields
\begin{equation}
\label{eq:Psc_st}
  \Psc=\frac{m_ec^2}{\l1}\frac{(\epKN^{\rm acc})^{\als+1}}{\als+1}
       \frac{n_0}{2}\left(e^{\vp/a}+e^{-\vp/a}\right).
\end{equation}
The medium accelerates according to the momentum equation~(\ref{eq:mom}).
With $\gamma\approx 1$ and $\rho\approx \rho_0$ this equation reads
\begin{equation}
\label{eq:mom_st}
  \rho_0c^2\frac{\dd\beta}{\dd\vp}=\Ppm+\Psc.
\end{equation}
Substituting (\ref{eq:Ppm_st}) and (\ref{eq:Psc_st}) and 
integrating for $\beta$ we get
\begin{eqnarray}
\nonumber
   \beta=\frac{m_ea}{2\mu_em_p\l1}
        \left[\frac{\ph(\alh-1)\epKN^{\alh-\als-1}a}{(\alh-\als-1)\l1}
              \left(e^{\vp/a}+e^{-\vp/a}-2\right)\right.\\ \left. 
              +\frac{(\epKN^{\rm acc})^{\als+1}}{\als+1}
               \left(e^{\vp/a}-e^{-\vp/a}\right)
        \right].\;\;\;{\rm \hspace*{1cm}}
\label{eq:acc}
\end{eqnarray}
Here we used $\rho_0/n_0=\mu_em_p$ (see eq.~\ref{eq:ion}). 
The non-relativistic zone ends when $\beta$ reaches $\sim 0.5$. 
Equating $\beta=0.5$ and neglecting the 
decaying exponential we get the acceleration length
(with $\als=0$, $\epKN=0.4$, and $\epKN^{\rm acc}=0.7$),
\begin{equation}
\label{eq:ac}
   \frac{\vpacc}{a}\approx\ln\frac{(\mu_em_p/m_e)(\l1/a)}
   {\frac{\ph(\alh-1)\epKN^{\alh-1}}{(\alh-1)}(a/\l1)+0.7}\approx 5+\ln\mu_e.
\end{equation}
Hence $\xiacc\approx 5\xiload$ at $\mu_e=1$, in full agreement with the 
numerical simulations (Fig.~1 and 3). As one can see from
equation~(\ref{eq:ac}), with $\mu_e=2$ the result changes only slightly,
$\xiacc\approx 5.7\xiload$. Note that $\xiload$ does not depend on $\mu_e$ 
at all. Hence the front structure is not sensitive to the chemical composition
of the ambient medium.

\subsection{Relativistic zone ($\beta\rightarrow 1$)}

At $\vp>\vpacc$ the medium continues to accelerate relativistically. 
Then $\epKN$ grows (eq.~\ref{eq:epKN}) and exceeds unity. 
The integral over $\vpp$ in equations~(\ref{eq:ndot1}) and (\ref{eq:Ppm1}) 
is now taken over two regions: $0<\vpp<\vp_1$ where $\epmax(\vpp)<1$ and 
$\vp_1<\vpp<\vp$ where $\epmax(\vpp)>1$. The boundary $\vp_1$ is defined by 
condition $\epmax=1$,
\begin{equation}
\label{eq:vp1}
 \vp-\vp_1=\frac{\l1}{\phi}\left(\frac{1+\beta_1}{1-\beta_1}\right)^{\alh}.
\end{equation}
This is an implicit equation for $\vp_1$ where $\beta_1=\beta(\vp_1)$.
One can show that $\vpacc<\vp_1\ll\vp$ when $\gamma(\vp)\gg 1$. From 
equation~(\ref{eq:ndot1}) we then find
\begin{eqnarray}
  \dot{n}_+=\frac{\phi c}{\lambda_1^2}\int_0^\vp \dd\vp^\prime\,
    \frac{n^*(\vpp)}{1+\bp} 
     \left(\frac{1-\bp}{1+\bp}\right)^{\alh} Q(\vpp),\hspace*{0.9cm} \\
  Q(\vpp)=
  \left\{\begin{array}{ll}
         \frac{\epmax^{\alh-\als}}{\alh-\als}, & \vpp<\vp_1,\\
         \frac{1}{\alh-\als}+\ln\min\{\epmax,\epKN\}, & \vpp>\vp_1.\\
  \end{array}\right.
\nonumber
\end{eqnarray}
The integral peaks at $\vpp\sim \vp_1$ 
(where $\epmin\ll \epmax\sim 1$ and we therefore set $\epmin\approx 0$ in the 
expression for $Q$). Denote the integrand as $S$ and evaluate the integral as 
$\int\dd\vpp S\approx \zeta\vp_1S(\vp_1)$.
We have $S\propto n^*$ at $\vpp<\vp_1$ and a steep decline 
$S\propto n^*{\gamma^\prime}^{2\alh}$ at $\vpp>\vp_1$, hence
$\zeta\approx(\dd\ln n^*/\dd\ln\vp+1)^{-1}$. From the numerical results 
we see that $\zeta\approx 1/3$. Then we get
\begin{eqnarray}
 \dot{n}_+(\vp)\approx
   \frac{\zeta c n_1^*\vp_1}{(\alh-\als)\l1(\vp-\vp_1)}.
\label{eq:ndot_rel}
\end{eqnarray}
This formula gives a reasonable approximation to $\dnp$ at $\vp>\vpacc$
(see Fig.~3). 

The approximation $\epmin\ll 1$ used in the derivation of
equation~(\ref{eq:ndot_rel}) breaks when $\epmin(\vp_1)$ approaches unity i.e. 
$\epthr$ for scattered photons with $\ep=1$ approaches $\epbr$. 
We define a characteristic $\vpsc$ such that $\epmin(\vp_1=\vpsc)=1/2$
(i.e. $\epthr=\epbr/2$). The velocity of the scattering medium at $\vpsc$
is given by (see eq.~\ref{eq:epmin})
\begin{equation}
\label{eq:gsc}
   \frac{1+\beta_{\rm sc}}{1-\beta_{\rm sc}}=\frac{\epbr}{4},
  \qquad \left(\gsc\approx \frac{\sqrt{\epbr}}{4} \quad {\rm if}\;\, 
              \epbr>100 \right).
\end{equation}
At $\vp_1>\vpsc$ the scattered peak $\ep\sim 1$ is not absorbed at any $\vp$.
The corresponding cut off in pair loading appears at 
\begin{equation}
\label{eq:vpbr}
 \vpbr=\frac{\l1}{\phi}\left(\frac{\epbr}{4}\right)^{\alh}
   \approx \vpacc\left(\frac{\epbr}{4}\right)^{\alh}.
\end{equation}

We conclude that (1) pair loading at any $\vp$ is sensitive to the medium
dynamics at $\vp<\vpsc$ only and (2) the extension of the pair loading zone
is limited by finite $\epbr$. 

The simple qualitative picture of pair loading in the relativistic zone
is as follows. 
The scattering of photons with $\ep\sim 1$ makes dominant
contribution to $\dnp$ at any $\vp<\vpbr$. Photons scattered at a given 
$\vp_1>\vpacc$ get absorbed at $\vp=\vp_1+\lgg\approx\lgg$ where 
\begin{equation}
  \lgg\sim\vpacc\left(\frac{1+\beta_1}{1-\beta_1}\right)^{\alh},
  \qquad \beta_1>0.5.
\end{equation}
The scattering in a narrow interval $\xiacc<\xi<3\xiacc$ controls pair 
loading in the whole relativistic zone $\xiacc<\xi<10^8$. Unfortunately, 
we do not have any simple analytical solution at $\xiacc<\xi<3\xiacc$.
The empirical formulae 
\begin{eqnarray}
\label{eq:appr1}
   \gamma=
  \left\{\begin{array}{ll}
    \left(\frac{\xi}{\xiacc}\right)^3, & \xiacc<\xi<3\xiacc, \\ 
  3\sqrt{3}\left(\frac{\xi}{\xiacc}\right)^{3/2}, & \;\; \xi>3\xiacc,\\
  \end{array}\right.\\
\label{eq:appr2}
   \frac{n^*}{n^*_{\rm acc}}=
  \left\{\begin{array}{ll}
    \left(\frac{\xi}{\xiacc}\right)^2, & \xiacc<\xi<3\xiacc, \\
  3\left(\frac{\xi}{\xiacc}\right), & \;\; \xi>3\xiacc,\\
  \end{array}\right.
\end{eqnarray}
fit well the numerical results for both $\alh=1.5$ and $\alh=2$ (see
Fig.~1 and 3).  Here $n^*_{\rm acc}\approx 
0.5\mu_e e^5n_0\approx 74\mu_e n_0$ is known from \S~5.2 (eqs.~\ref{eq:stat} 
and \ref{eq:ac}).

\subsection{Heating by pair loading}

We now give estimates for the mean energy and momentum
of the injected pairs, first in the lab frame ($e_\pm$, $p_\pm$) and then
in the rest frame of the medium ($\ginj$, $\pinj$). The estimates 
highlight the role of $e^\pm$ in the heating and acceleration of the medium. 

In the lab frame, the energy and momentum of a created $e^\pm$ pair is 
dominated by the absorbed primary (collimated) photon,
\begin{equation}
 \frac{e_\pm}{m_ec^2}\approx\frac{p_\pm}{m_ec}\approx\epabs=\chi\epthr,
\end{equation}
where we used equations~(\ref{eq:gpm}) and (\ref{eq:chi}) of Appendix. 
This expression can be further averaged over the spectrum of scattered photons.
The averaged values can be written as $\bar{e}_\pm/c\approx \bar{p}_\pm
=\Ppm/\dnp$. 

\centerline{ \epsfxsize=9.0cm {\epsfbox{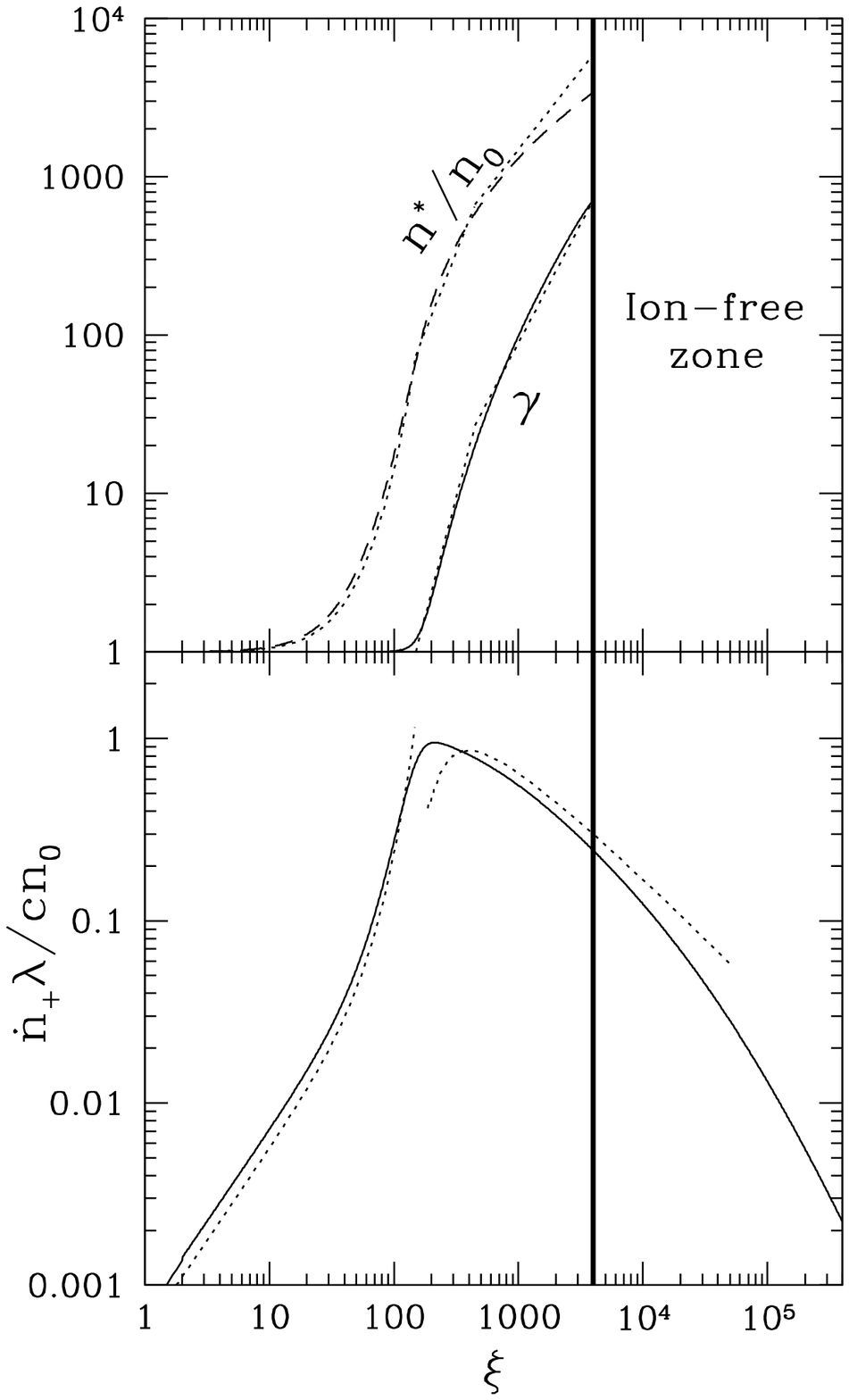}} }
\figcaption{Comparison of the analytical model ({\it dotted curves}) with the 
exact numerical results for $\alpha_2=2$ and $\epbr=10^2$. 
Here $\xiload\approx 30$ (see eq.~\ref{eq:stat}) and $\xiacc=5\xiload$
(see eq.~\ref{eq:ac}). The break in 
$\dnp$ is at $\xibr\sim 5\times 10^4$. The boundary of the ion-free zone
$\xi_c$ is shown by {\it solid vertical line}. In this example,
$\xi_c=4\times 10^3$ is chosen close to its typical value (see \S~7).}
\bigskip

In the non-relativistic zone of the front the created $e^\pm$ have 
initial Lorentz factors $\ginj=1+e_\pm/2m_ec^2$. Upon collectivization
they push the medium forward and heat it. Using equations~(\ref{eq:ndot1}) 
and (\ref{eq:Ppm1}) we have 
\begin{equation}
\label{eq:hh}
  \ginj-1=\frac{\Ppm}{2\dnp m_ec}
    =\frac{(1+\alh^{-1})^{5/3}(\alh-\als)}{(\alh-\als-1)\epKN}.
\end{equation}
For hard spectra with finite $\epbr$ this equation overestimates 
$\ginj$ (see discussion after eq.~\ref{eq:Ppm_st}); e.g. for $\alh=1.5$ and
$\epbr=10^2$ the actual $\ginj\approx 9$ (Fig.~1) while 
equation~(\ref{eq:hh}) gives $\ginj\approx 19$. 

In the relativistic zone, $\xi\gg\xiacc$, we have 
\begin{equation}
 \frac{e_\pm}{2m_ec^2\gamma}=\frac{\chi\epthr}{2\gamma}
  \approx \frac{1}{\gamma}\left(\frac{\phi\vp}{\l1}\right)^{1/\alh}
  \sim \frac{0.1\xi^{1/\alh}}{\gamma}\ll 1,
\end{equation}
i.e. the Lorentz factor of created pairs is smaller than that of the medium. 
It implies that the $e^\pm$ loading tries to decelerate the medium. Same 
effect can be viewed from the medium rest frame (see eq.~\ref{eq:ppp} below).

To find the Lorentz factor of created $e^\pm$ in the medium rest frame,
we use equations~(\ref{eq:gpm}) and (\ref{eq:ginj}) of Appendix and
substitute $\mu=\beta_1$, $\epsc=(1+\beta_1)^{-1}$, and 
$\epthr=2(1+\beta_1)/(1-\beta_1)$ (cf. \S\S~5.1 and 5.3). This yields
\begin{eqnarray}
\nonumber
 \ginj
  =\frac{\gamma}{2}\left[\epsc(1-\beta\beta_1)+\chi\epthr(1-\beta)\right] \\
  \approx\frac{\gamma}{\epthr}+\frac{\chi}{4}\frac{\epthr}{\gamma}
   \approx\frac{\gamma}{\epthr}\gg 1,
\label{eq:ginj_rel}
\end{eqnarray}
(we used $\gamma\gg\gamma_1$). Here we keep the $\epsc$ term and 
neglect the $\epabs$ term: in the rest frame, the scattered photon
[blueshifted as $\gamma(2\gamma_1^2)^{-1}>1$] is more 
energetic than the primary 
collimated photon [redshifted as $(2\gamma)^{-1}$].
In a similar way we evaluate momentum per injected particle
in the medium rest frame,
\begin{equation}
\label{eq:ppp}
  \pinj\sim
\left\{\begin{array}{ll}
  \epthr, & \xi<\xiacc, \\
  -\gamma/\epthr, & \xi\gg\xiacc. \\
\end{array}\right.
\end{equation}
The $e^\pm$ loading assists the medium acceleration only as long as 
$\pinj>0$. At $\xi\simgt\xiacc$, $\pinj$ changes sign. At this point,
$\epthr/\gamma\sim 1$ and $\ginj\sim (\epthr/\gamma)+(\gamma/\epthr)$ reaches
its minimum $\sim 2$. The minimum of $\ginj$ is observed in Figure~1.

Note that collectivization of the relatively slow $e^\pm$ loaded at 
$\xi\gg\xiacc$ implies their fast acceleration (owing to e.g. stream 
instability, with the $e^\pm$ stream being directed backward in the medium 
rest frame). In the case of partial collectivization (cf. \S~2.2) the 
isotropized $e^\pm$ acquire Lorentz factor 
$\gamma_e\sim 2\gamma m_ec^2/e_\pm\gg 1$ found from 
equation~(\ref{eq:ginj_rel}) and then they are cooled by radiation. 
The cooling is dominated by photons with $\ep\sim 1$ and occurs in Thomson 
regime. When viewed from the lab frame, the hot $e^\pm$ move forward with 
Lorentz factors between $\gamma_{\rm min}=e_\pm/2m_ec^2$ and
$\gamma_{\rm max}=2\gamma^2m_ec^2/e_\pm$.
They scatter photons with energy $\epsc\approx \ep/2\sim 1/2$ 
(eq.~\ref{eq:toy2}) through angles 
$1-\mu\simlt (e_\pm/m_ec^2)^{-2}=(\chi\epthr)^{-2}$. 
If the scattered radiation produced by this cooling process could interact 
with the primary beam we would have secondary pair production. This 
however requires $2/[\epsc(1-\mu)]<\epbr$ which in turn
requires $\epthr<\sqrt{\epbr}/4\chi$ for the primary injection event.
This condition is not satisfied at $\epbr<10^3$ and the 
secondary pair production does not occur.


\section{Backreaction on GRB}

\subsection{Scattering}

The observed GRB can be affected by scattering in the circumburst medium if
the Thomson optical depth of the medium is substantial.
In all calculations we assumed that the medium remains optically thin even 
after $e^\pm$ loading. We now address this assumption.

Consider a radius $R$ and let the ambient medium have an initial optical 
depth $\tau_R=n_0(R)\sT R$ at this radius. In a medium with constant electron
density $n_0(R)=const$,
\begin{equation}
\label{eq:tau_ISM}
  \tau_R=7\times 10^{-10}R_{15}n_0. 
\end{equation}
In a steady wind with mass loss $\dM$ and velocity $w$,
\begin{equation}
\label{eq:tau}
  \tau_R=3\times 10^{-4}R_{15}^{-1}\frac{\dM_{21}}{\mu_ew_8}. 
\end{equation}
The pair loading in the radiation front increases $\tau_R$.
The optical depth seen by the GRB photons at given $\vp$ is
\begin{equation}
\label{eq:tau_load}
  \tau_R^*=\tau_R\frac{n^*(\vp)}{n_0}.
\end{equation}
Suppose $\tau_R^*$ reaches unity at some $\vpcr$. Radiation scatters here
off the medium with $\gcr=\gamma(\vpcr)$ and acquires a new collimation angle
$\theta\sim \gcr^{-1}$. This decollimation is not crucial if $\gcr$ is
sufficiently large, $\gcr>\gamma_{\rm min}\approx 10^2$.
Using the solution for $n^*/n_0$ and $\gamma$ (Fig.~1 and 3) one finds that 
the condition $\tau_R^*<1$ at $\gamma=100$ reads
\begin{equation}
\label{eq:tau_cr}
  \tau_R<\taucr\approx 10^{-3}.
\end{equation}
This constraint slightly changes with a different choice of $\gamma_{\rm min}$.

Even assuming that the burst is sufficiently short (so that it decouples
promptly from the ejecta, cf. case 2 in \S~7.2) the whole burst can overtake 
the medium it sweeps only at $R=R_c$ (eq.~\ref{eq:R_c}). 
The condition~(\ref{eq:tau_cr}) is therefore required at $R>R_c$.
In the ISM case ($n_0=const\sim 1$~cm$^{-3}$) this condition is satisfied 
for any reasonable parameters.
In the wind case the condition $\tau_R(R_c)<\taucr$ reads (using 
eq.~\ref{eq:R_c}) 
\begin{equation}
\label{eq:limit}
   \frac{\dM_{21}}{\mu_ew_8}<7E_{53}^{3/7}t_b^{1/7}, 
  \qquad t_b=\frac{\Delta}{c}.
\end{equation}

\subsection{$\gamma-\gamma$ absorption and the high-energy break}

In previous sections we assumed {\it a priori} a high-energy break in the
primary radiation spectrum at some $\epbr$.
In a self-consistent situation, $\epbr$ is determined by $\gamma-\gamma$ 
absorption of the primary $\gamma$-rays by the scattered radiation field.
One can evaluate $\epbr$ in a simple way. 
At given $\vp$ the primary photons $\ep\sim\epthr$ are absorbed with 
rate $\dnp(\vp)$. The number of absorbed photons during time $R/c$ cannot 
exceed their initial number. This condition reads
$\dnp R/c<(F_1/m_ec^3)\epthr^{-\alh-1}$ and gives an upper limit on $\epthr$
i.e. the self-consistent $\epbr$. Using equations (\ref{eq:ndot_rel}),
(\ref{eq:vp1}), and $\epthr=2(1+\beta_1)/(1-\beta_1)$, we get
(omitting a numerical factor $\sim 1$)
\begin{equation}
\label{eq:br}
   \epthr<\frac{\xiacc}{\xi_1}\frac{1}{n^*_1R\sT}
   =\frac{\xiacc}{\xi_1}\frac{n_0}{n^*_1}\frac{1}{\tau_R}.
\end{equation}
Hence $\epbr>1$ if $\tau_R(n^*_{\rm acc}/n_0)<1$, i.e. the main radiation 
$\ep\sim 1$ is not self-absorbed after scattering if
\begin{equation}
\label{eq:tau_abs}
  \tau_R<\frac{n_0}{n^*_{\rm acc}}\approx 10^{-2}.
\end{equation}
This condition is weaker than the transparency condition~(\ref{eq:tau_cr}). 
To find $\epbr$ at $\tau_R<10^{-2}$ we need to solve the 
inequality~(\ref{eq:br}) which is implicit since 
$\epthr$ is a function of $\xi_1$.
The solution gives the maximum $\xi_1^{\rm max}$ and the corresponding
$\epthr^{\rm max}=\epbr$. 
At $\tau_R\ll 10^{-2}$ we have $\epbr\gg 1$; then
$\epthr\approx 8\gamma_1^2=8(\xi_1/\xiacc)^6$ and
$n^*_1=74\mu_e n_0(\xi_1/\xiacc)^2$ (using eqs.~\ref{eq:appr1} and 
\ref{eq:appr2}). We thus find 
$\xi^{\rm max}_1/\xiacc=(600\mu_e\tau_R)^{-1/9}$ and
\begin{equation}
\label{eq:break}
  \epbr\approx 0.1(\mu_e\tau_R)^{-2/3}, \qquad \tau_R\ll 10^{-2}.
\end{equation}
The break appears in the GRB if $\vpbr<\Delta$
where $\vpbr$ is given by equation~(\ref{eq:vpbr}). We have from
equation~(\ref{eq:vpbr})
(using $\vpacc\approx 120\lambda$ and eq.~\ref{eq:lam})
\begin{equation}
\label{eq:vpbr_tau}
  \vpbr\approx 6\times 10^8\frac{R_{15}^2}{L_{53}}
               \left(\frac{\epbr}{4}\right)^{\alh}{\rm cm}
 \approx\frac{6\times 10^8R_{15}^2L_{53}^{-1}}{40^{\alh}
     (\mu_e\tau_R)^{2\alh/3}} {\rm~cm}.
\end{equation}
The condition $\vpbr<\Delta$ reads
\begin{equation}
\label{eq:br_cond}
  \frac{\vpbr}{\Delta}\approx\frac{2\times 10^{-2}R_{15}^2E_{53}^{-1}}
    {40^{\alh}(\mu_e\tau_R)^{2\alh/3}}<1.
\end{equation}

If the ambient medium is ISM with the optical depth~(\ref{eq:tau_ISM}) 
then the condition~(\ref{eq:br_cond}) is not satisfied outside the emission 
radius of the GRB and hence the GRB-medium interaction does not produce any 
break in the GRB spectrum. 

If the ambient medium is a wind with optical depth~(\ref{eq:tau}) then 
the condition~(\ref{eq:br_cond}) is satisfied at radii $R<\Rgg$ where
\begin{equation}
\label{eq:Rgg}
  \Rgg\approx10^{15}\left(\frac{50E_{53}}{5.6^{\alh}\mu_e}
               \right)^{3/(6+2\alh)}
       \left(\frac{\dM_{21}}{w_8}\right)^{2\alh/(6+2\alh)}{\rm cm}.
\end{equation}
For instance, with $\alh=2$,
$\Rgg\approx 10^{15}E_{53}^{3/10}(\dM_{21}/w_8)^{2/5}$cm. 
It can be well outside the emission 
radius $\Rem$ and cause a break in the GRB spectrum. 

We now derive $\epbr$ expected in the 
massive progenitor scenario. We substitute $\tau_R$ from 
equation~(\ref{eq:tau}) into equations~(\ref{eq:break}) and 
(\ref{eq:vpbr_tau}) and get
\begin{equation}
\label{eq:epbr_R}
   \epbr\approx 22R_{15}^{2/3}\left(\frac{w_8}{\dM_{21}}\right)^{2/3}.
\end{equation}
\begin{equation}
\label{eq:vpbr_R}
  \vpbr\approx 6\times 10^8\times 5.6^{\alh}\frac{R_{15}^{(6+2\alh)/3}}
    {L_{53}} \left(\frac{w_8}{\dM_{21}}\right)^{2\alh/3}{\rm cm}.
\end{equation}
The lowest $\epbr$ is produced at small $R$ in the leading portion of the 
radiation front $\vpbr\ll\Delta$. With increasing $R$, $\vpbr$ and $\epbr$ 
grow. At $R=\Rgg$ the $\vpbr$ reaches $\Delta$ and ends up outside the front,
and hence at $R>\Rgg$ the $\gamma$-radiation does not change. 
From equations~(\ref{eq:epbr_R}) and (\ref{eq:vpbr_R}) one gets 
$\epbr\propto\vpbr^{1/(\alh+3)}$. 
One can show that during the spectrum shaping at $R<\Rgg$ the dependence 
$\epbr\propto\vp^{1/(\alh+3)}$ is established at $\vp<\vpbr$. At 
$\vpbr<\vp<\Delta$ a break is temporarily formed at a higher energy
$\epbr\propto \vp^{1/(\alh+7/6)}$ (one finds 
the ``extinction'' zone $\vp$ of primary photons $\ep$ by equating their 
number to the number of scattered photons that have $\epthr=\ep$).
At $R>\Rgg$, the whole front has $\epbr\propto \vp^{1/(\alh+3)}$.

A distant observer will see first the leading portion and then deeper layers 
of the $\gamma$-ray pulse (the observer time is $\tobs=\vp/c$).
Eliminating $R$ from equations~(\ref{eq:epbr_R}) and (\ref{eq:vpbr_R})
we find
\begin{equation}
\label{eq:break_wind}
   \epbr(\tobs)=(30\pm 10)
 \left(\tobs L_{53}\right)^{1/(\alh+3)}
 \left(\frac{w_8}{\dM_{21}}\right)^{2/(\alh+3)}.
\end{equation}
Here 30 corresponds to $\alh=1.5$, and the upper and lower values correspond
to $\alh=1$ and 2.5, respectively.

The equation~(\ref{eq:break_wind}) has been derived assuming an idealized 
GRB with $F(\vp)=const$, while the observed GRBs are highly variable.
The GRB variability did not play a big role to the structure of the front 
(see footnote in \S~4); in contrast, $\epbr(\vp)$ is sensitive to variability. 
The scattered radiation that shapes $\epbr(\vp)$ comes from a small leading 
portion of the GRB fluence. Therefore $\epbr$ is sensitive to the ratio of 
the flux at given $\vp$ to that in the leading portion. The higher $F(\vp)$ 
the less depleted its high-energy part is, which results in a higher 
$\epbr(\vp)$. Hence a strong positive correlation between the instantaneous
$F$ and $\epbr$ should be observed in variable GRBs.


\section{Evolution of the radiation front}

In this section we study the front evolution with radius.
We assume the internal scenario of GRB production (\S~1) and start from 
radius $\Rem$ where the emitted $\gamma$-ray pulse begins to decouple from 
the ejecta. It happens when the ejecta becomes transparent,
\begin{equation}
\label{eq:Rem}
  \Rem=KR_*\approx 6\times 10^{13}Kt_b^{-1}{\Eej}_{53}{\Gej}_2^{-3}{\rm ~cm}. 
\end{equation}
Here $R_*$ is the radius of ``barion'' transparency and $K>1$ describes a 
possible increase of the transparency radius owing to pair creation 
inside the ejecta; $\Eej=\Gej\Mej c^2$ is the energy of the ejecta.
Note that $K\gg 1$ would require: (1)~a substantial fraction of the emitted 
energy is in very hard $\gamma$-rays (above the threshold for pair creation, 
$\ep>\Gej$), and 
(2)~the hard $\gamma$-rays are emitted at a high rate at radii $R\gg R_*$
(otherwise $e^\pm$ production stops, pairs immediately annihilate 
to optical depth $\sim 1$ and the ejecta becomes transparent on time-scale
$R/c$ because of side expansion). It is unclear whether the two conditions 
are satisfied.
The observed strong variations in many GRBs on time-scales 
$\delta t< 0.1$~s suggest that in many cases $\Rem<2\Gej^2c\delta t=
6\times 10^{13}{\Gej}_2^2(\delta t/0.1)$~cm.

The thickness of the radiation pulse is equal to that of the ejecta, 
$\Delta\approx\Dej$. Radiation is initially 
collimated within angle $\theta=\Gej^{-1}$ and moves inside the ejecta.
At $R>\Rem$ the radiation gets more collimated, 
$\theta=\Gej^{-1}(R/\Rem)^{-1}$, and
gradually overtakes the ejecta with relative velocity $\approx(1-\bej)c$.
The thickness of the radiation pulse emerging ahead of the ejecta and 
interacting with the ambient medium is growing,
$\Delta_i(R)=c(1-\bej)(R/c)=R/2\Gej^2$. When $\Delta_i(R)$ reaches $\Dej$,
the whole pulse has left the ejecta.
The corresponding $\xi$-coordinate of the back boundary of the interacting 
pulse $\xiD$ is 
\begin{equation}
\label{eq:xiD}
  \xiD(R)=
  \left\{\begin{array}{ll}
 \frac{\Delta_i}{\lambda}\approx
 1.1\times 
  10^4R_{15}^{-1}L_{53}{\Gej}_2^{-2}, &  \frac{R}{2\Gej^2}<\Dej, \\
 \frac{\Dej}{\lambda}\approx 6.5\times 10^3R_{15}^{-2}E_{53}, & 
   \frac{R}{2\Gej^2}>\Dej. \\  
  \end{array}\right.
\end{equation}

\subsection{$\Rem<R<\Rsat$. Saturated surfing}

The pulse-medium interaction starts at $R\simgt\Rem$ with a very high 
$\xiD\sim 10^6$. The medium entering the pulse accelerates to 
the equilibrium Lorentz factor
\begin{equation}
\gsat(R)=\Gej\frac{R}{\Rem} 
\end{equation}
at $\xi_c\sim 10^3\ll\xiD$ and surfs the pulse. 
The acceleration time is $\sim(\xi_c\lambda/c)\gsat^2< R/c$.
Note that primary radiation is mixed in the front on scale 
$\delta\vp_{\rm mix}\sim \gsat^{-2}R$ because the photons have a finite 
angular dispersion $\theta\sim \gsat^{-1}$. The
Lagrangian coordinate $\vp$ is well defined only on scales 
$\delta\vp>\vp_{\rm mix}$ (on such scales the radiation can be assumed
perfectly collimated with radial velocity $c$). Therefore, the 
$\xi$-location of the medium in the front is defined with uncertainty 
$\ximix=\vp_{\rm mix}/\lambda\sim \xiD(R/\Rem)^{-2}$ which exceeds $\xi_c$
at small $R$ where $\xi_c/\xiD<(\Rem/R)^2$. 

The equilibrium Lorentz factor $\gsat(R)$ grows with radius. 
Correspondingly $\xi_c$ 
[the value of $\xi$ where $\gamma(\xi)$ reaches $\gsat$, see Fig.~1]
grows and reaches $\sim 10^{4}$ at $R\sim 10^{14}$~cm.
At $R=\Rsat$,
\begin{equation}
\label{eq:Rsat}
  \Rsat\approx 1.3\times 10^{14}{\Gej}_2^{-3/4}L_{53}^{1/4}
   {\Rem}_{13}^{3/4}{\rm ~cm},
\end{equation}
$\gsat$ exceeds $\gmax$ given by equation~(\ref{eq:gmax}). 
Then the medium cannot accelerate to $\gsat$ on time $R/c$ and the 
saturated stage ends.

\subsection{$\Rsat<R<\Rgap$. Unsaturated surfing: caustic}

Now $\xi_c$ and $\gc$ are determined by the condition 
$(\lambda\xi_c/c)\gc^2\approx R/c$ (the time of acceleration to $\gc$ is 
about $R/c$). Using equation~(\ref{eq:appr1})  we get at $\gc>27$,
\begin{eqnarray}
\label{eq:xic}
  \frac{\xi_c}{\xiacc}\approx\frac{53}{\xiacc^{1/4}}L_{53}^{1/4}R_{15}^{-1/4},
\\
  \gc\approx \frac{2.0\times 10^3}{\xiacc^{3/8}}L_{53}^{3/8}R_{15}^{-3/8}.
\label{eq:gc}
\end{eqnarray}
When $R$ grows from $10^{14}$cm to $10^{16}$cm, $\xi_c$ decreases slowly
from $\xi_c\approx 30\xiacc$ to $\xi_c\approx 10\xiacc$. Correspondingly,
$\gc$ decreases from $\approx 10^3$ to $\approx 140$. Hereafter we substitute 
in all estimates $\xiacc=120$ keeping in mind the typical $\alh=1.5$; for 
$\alh=2$ there is a slight change: $\xiacc=150$ (\S~5.2).

The new material trapped at given $R$ comes to $\xi_c$ with smaller 
$\gamma$ compared to that of the already accumulated material in the front. 
This results in ``overshooting'' and implies appearance of a caustic.
The overshooting  can be seen e.g. in the $\vp$-coordinate:
the accumulated material has $\vp_c^{\rm old}=\xi_c\lambda$ and the 
newly trapped material comes to $\vp_c^{\rm new}=R/\gc$. Hence,
$\vp_c^{\rm new}/\vp_c^{\rm old}\propto \xi_c/\gamma_c$. 
With decreasing $\xi_c$, the condition for overshooting reads 
$\dd(\xi/\gamma)/\dd\xi<0$ at $\xi=\xi_c$, or
\begin{equation}
  \left[\frac{\dd\ln\gamma}{\dd\ln\xi}\right]_{\xi=\xi_c}>1,
\end{equation}
which is satisfied (see eq.~\ref{eq:appr1}).
The caustic results in a shock. If the shock is radiative
(which may be the case since the material is pair-dominated and the 
Compton cooling is very efficient) then the shocked matter piles up in 
a thin shell.

When the caustic appears, the accumulated ion material 
begins to decelerate and the $e^\pm$ stream loaded behind $\xi_c$
hits the ion medium. One can show that the momentum of the $e^\pm$
stream exceeds the momentum of the accumulated ions
and this ``reverse'' shock should be strong. 
For simplicity, we will neglect the impact of the $e^\pm$ stream  on the 
surfing medium (the inclusion of this effect will slightly increase the 
radius $\Rgap$ derived below).

The medium surfs the pulse with $\gamma\approx\gc$ until $\xiD$ reaches 
$\xi_c$. This happens at some radius $R_c$.
We now evaluate $R_c$ in two possible cases.

1. $R_c<2\Gej^2\Dej$. --- The ejecta catches up with the 
surfing medium before the whole $\gamma$-ray pulse leaves the ejecta.
Then $\xi_c=\xiD$ gives (using eqs.~\ref{eq:xiD} and \ref{eq:xic})
\begin{equation}
\label{eq:R_c_long}
  R_c=\frac{1.2\times 10^{18}}{\xiacc}\frac{L_{53}}{{\Gej}_2^{8/3}}{\rm~cm}
     \approx 10^{16}\frac{L_{53}}{{\Gej}_2^{8/3}}{\rm~cm}.
\end{equation}
At $R=R_c$ we also have $\gc\approx\Gej$ i.e. the ejecta touches the medium 
with a small relative Lorentz factor and starts to decelerate. The 
gap between the surfing medium and the ejecta disappears at this moment; 
we thus have $\Rgap=R_c$.

The assumed condition $R_c<2\Gej^2\Dej$ requires $t_b>t_b^*$, 
\begin{equation}
\label{eq:tb}
  t_b^*=\frac{45}{\xiacc^{1/2}}\frac{E_{53}^{1/2}}{{\Gej}_2^{7/3}}{\rm~s}
   \approx 4\frac{E_{53}^{1/2}}{{\Gej}_2^{7/3}}{\rm~s},
\end{equation}
where we used $t_b=\Dej/c=E/L$. This condition implies
\begin{equation}
\label{eq:Rgap_long}
  \Rgap=R_c<\frac{3\times 10^{16}}{\xiacc^{1/2}}
            \frac{E_{53}^{1/2}}{{\Gej}_2^{1/3}}{\rm ~cm}
    \approx 3\times 10^{15}\frac{E_{53}^{1/2}}{{\Gej}_2^{1/3}}{\rm ~cm}.
\end{equation}

2. $R_c>2\Gej^2\Dej$. --- The whole $\gamma$-ray pulse 
leaves the ejecta before they reach $R_c$. Then $\xi_c=\xiD$ at
\begin{equation}
\label{eq:R_c}
  R_c=\frac{1.6\times 10^{16}}{\xiacc^{3/7}}E_{53}^{3/7}t_b^{1/7}{\rm ~cm}
   \approx 2\times 10^{15}E_{53}^{3/7}t_b^{1/7}{\rm ~cm}.
\end{equation}
The value of $\gc(R_c)$ now differs from $\Gej$,
\begin{equation}
   \gc=\frac{7.1\times 10^2}{\xiacc^{3/14}}E_{53}^{3/14}t_b^{-3/7}.
\end{equation}
The condition $R_c>2\Gej^2\Dej$ (which is equivalent to $t_b<t_b^*$, cf.
eq.~\ref{eq:tb}) implies that $\gc>\Gej$ and hence the gap still
exists at $R=R_c$. The gap disappears only when $\gamma(\xiD)\approx\Gej$ 
(then the ejecta catches up with the surfing medium). This condition yields
\begin{equation}
\label{eq:Rgap}
   \Rgap=\frac{3.0\times 10^{16}}{\xiacc^{1/2}}
  \frac{E_{53}^{1/2}}{{\Gej}_2^{1/3}}{\rm ~cm}
        \approx 2.7\times 10^{15}\frac{E_{53}^{1/2}}{{\Gej}_2^{1/3}}{\rm ~cm}.
\end{equation}

We call case~1 as ``long-burst'' regime ($t_b>t_b^*$) and case~2 
as ``short-burst'' regime ($t_b<t_b^*$).
Note that $\Rgap$ is smaller in case~1 (compare eqs.~\ref{eq:Rgap} and
\ref{eq:Rgap_long}). The gap may be not opened if the GRB emission radius
is much increased by $e^\pm$ inside the ejecta (cf. 
eq.~\ref{eq:Rem}) and exceeds $\Rgap$;
$\Rem>\Rgap$ would require $K>170 {\Gej}_2^{-2/3}(E/\Eej)$.

Hereafter in this paper we focus on the short-burst regime. Then the energy
of the $\gamma$-ray pulse interacting with the ambient medium is constant 
at $R>R_c$ and the simple scaling $\xiD\propto R^{-2}$ holds.
A simplified blast wave model can be constructed for short bursts  (as done 
in \S~8). The extension to long GRBs is straightforward, though it implies 
additional technical details which we defer to a future paper.

\subsection{$\Rgap<R<\Racc$. Preaccelerated pair-rich medium}

In this range of radii, $\xiD$ decreases from $\xi_c(\Rgap)\approx 10^3$ 
to $\xiacc\approx 10^2$.
Correspondingly, $\gamma(\xiD)$ decreases from $\Gej$ to $\approx 1$.
When $\xiD<\xiacc$ the front cannot accelerate the medium
to relativistic velocities. The condition $\xiD=\xiacc$ therefore
defines the maximum radius where the relativistic preacceleration occurs,
\begin{equation}
  \Racc=\frac{R_\lambda}{\xiacc^{1/2}}
    \approx 7\times 10^{15} E_{53}^{1/2}{\rm ~cm}.
\label{eq:Racc}
\end{equation}

\subsection{$\Racc<R<\Rload$. Non-relativistic pair-rich medium}

At $R>\Racc$ the radiation front still loads the medium with a large number 
of pairs. At $R=\Racc$ ($\xiD=\xiacc$)
we have $n^*/n_0\approx 74\mu_e$ behind the front, and with increasing 
$R$ the pair loading decreases exponentially (see eq.~\ref{eq:stat}).
The pair loading ends at $R=\Rload$ (here $\xiD$ reaches $\xiload$ and
$n^*/n_0\sim 1$),
\begin{equation}
\label{eq:Rload}
  \Rload=\frac{R_\lambda}{\xiload^{1/2}}
  \approx 1.6\times 10^{16}E_{53}^{1/2}{\rm ~cm}.
\end{equation}
In \S~5 we showed that $\xiacc$ is related to $\xiload$ by a simple 
formula $\xiacc=(5+\ln\mu_e)\xiload$ which weakly depends on $\mu_e$
($1<\mu_e<2$). Hence, we have a relation
\begin{equation}
\label{eq:relation}
   \Rload=(5+\ln\mu_e)^{1/2}\Racc=(2.3\pm 0.1) \Racc.
\end{equation}

\subsection{$R>\Rload$. Front weakly affects the medium}

Here $\xiD<\xiload$ and both $e^\pm$ loading and preacceleration are shut down. 
The blast wave sweeps the normal pair-free medium which has $\gamma\approx 1$.


\section{Blast wave}

\subsection{Dynamics}

When the gap is closed, the ejecta drives a blast wave through 
the medium preaccelerated by the leading radiation front.
We will model the blast wave in a simplified way, as a thin shell 
sweeping the medium. This is a good approximation
to the exact hydrodynamic solution with forward and reverse shocks 
if the ejected shell is sufficiently thin, so that the reverse shock crosses 
$\Dej$ on time less than $R/c$ (e.g. Piran 1999)\footnote{
The standard blast wave model assumes the formation of collisionless shocks. 
Smolsky \& Usov (2000) developed and alternative model for the ejecta-medium 
interaction. The sweeping-shell approximation is useful in that case as well
as it deals with energy-momentum conservation only and gives a correct 
$\Rdec$.}.

The shell has initial mass $\Mej$ and Lorentz factor $\Gej$ and starts
to sweep the medium at $R=\Rgap$ (\S~7). At a radius $R>\Rgap$
the shell has mass $M>\Mej$ and Lorentz factor $\Gamma<\Gej$.
When it sweeps a mass element $\dd m$ that moves with Lorentz factor 
$\gamma$, $\Gamma$ decreases by $\dd\Gamma$ and energy $\dd \Ed$ is 
dissipated. The laws of energy and momentum conservation read
\begin{eqnarray}
\label{eq:laws}
  \Gamma M+\gamma\dd m=(\Gamma+\dd\Gamma)(M+\dd m+\dd\mheat), \\
\nonumber
  \Gamma\bh M +\gamma\beta\dd m
   =(\Gamma+\dd\Gamma)(\bh+\dd\bh)\hspace*{0.84cm}\\
  \times(M+\dd m+\dd\mheat).\hspace*{0.84cm}
\end{eqnarray}
Here $\bh=(1-1/\Gamma^2)^{1/2}$ is the shell velocity and 
$\dd\mheat=(\dd \Ed/c^2\Gamma)$ is the rest mass associated with
dissipated heat. The inertial mass $M$ includes the initial mass of
the ejecta $\Mej$, the swept mass $m(R)$, and the stored heat.
We will assume that a fraction $\eta$ of $\dd\mheat$ is radiated away
and the rest remains to increase the kinetic energy of the shell.
Then $\dd M=\dd m+(1-\eta)\dd\mheat$ and we get
dynamic equations,
\begin{eqnarray}
\label{eq:dyn1}
  M\frac{\dd\Gamma}{\dd m}=\Gamma^2\bh\gamma(\beta-\bh), \\
  \frac{\dd M}{\dd m}=\eta+(1-\eta)\Gamma\gamma(1-\bh\beta).
\label{eq:dyn2}
\end{eqnarray}
The radiated energy is  
\begin{equation}
\label{eq:Erad}
  \frac{\dd \Erad}{\dd m}=\eta\frac{\dd \Ed}{\dd m}
    =\eta c^2\Gamma\left[\Gamma\gamma(1-\bh\beta)-1\right].
\end{equation}
The swept mass is related to radius by $\dd m/\dd R=4\pi R^2 \rho_0$ where 
$\rho_0(R)$ is the medium density ahead of the radiation front.
Note that the increase of the medium mass by $e^\pm$ loading is 
fairly low at $R>\Rgap$ (\S~7). The front affects the blast wave dynamics 
mainly by increasing $\gamma$ of the ambient medium 
(and possibly by increasing the efficiency $\eta$ as a result of $e^\pm$
loading, see TM; here we assume $\eta=const$ for simplicity). The dynamic 
equations acquire the standard form if $\gamma=1$ (deceleration by static 
medium, see Piran 1999). 

The time interval between radiative preacceleration and subsequent sweeping 
by the ejecta shell is much smaller than $R/c$, and hence preacceleration 
can be treated locally at a given $R$. 
Indeed, the distance between the leading boundary of the radiation 
front and the blast is $\Delta_f=(1-\bh)R\approx R/2\Gamma^2$ and
the sweeping time is
$\tsw=(\Delta_f/c)2\gamma^2=(R/c)(\gamma/\Gamma)^2<R/c$ at $R>\Rgap$.
The Lorentz factor $\gamma$ at given $R$ is that 
found behind the radiation front, $\gamma=\gamma(\xiD)$ (\S\S~4 and 5).
The whole $\gamma$-ray pulse interacts with the ambient medium ahead of the
ejecta (we assume the short-burst regime, so that the whole pulse has left
the ejecta at $R<\Rgap$, cf. \S~7.2) and
\begin{equation}
\label{eq:x}
   \xiD=\left(\frac{R_\lambda}{R}\right)^2
   =\frac{\xiacc}{x^2}, \qquad 
   x\equiv\frac{R}{\Racc}.
\end{equation}
We will use the analytical formula~(\ref{eq:appr1}) for $\gamma(\xiD)$; then 
\begin{equation}
\label{eq:gam_x}
  \gamma(x)=
  \left\{\begin{array}{ll}
    1, & x>1, \\
    x^{-6}, & \frac{1}{\sqrt{3}}<x<1, \\
    3\sqrt{3}x^{-3}, & \xgap<x<\frac{1}{\sqrt{3}},
  \end{array}\right.
\end{equation}
where $\xgap=\sqrt{3}\Gej^{-1/3}\approx 0.3$ is found from $\gamma=\Gej$.

The characteristic mass of the problem is the ambient mass within the 
acceleration radius,
\begin{equation}
\label{eq:macc}
   \macc=\int_0^{\Racc}4\pi R^2\rho_0\dd R.
\end{equation}
The mass swept when the blast wave reaches a radius $x$ is
\begin{equation}
\label{eq:m_x}
   m(x)=\macc x^k,
\end{equation}
where $k=3$ for a constant-density medium and $k=1$ for a wind with 
constant $\dM$ and $w$.

At $\gamma\ll\Gamma$ equation~(\ref{eq:Erad}) yields
\begin{equation}
\label{eq:Ediss}
  \frac{\dd\Ed}{\dd m}\approx c^2\Gamma^2\frac{(1+\beta)}{\gamma}.
\end{equation}
Before the ejecta decelerates, $\Ed\ll\Eej=\Gej\Mej c^2$, we have
$\Gamma\approx \Gej$. Equation~(\ref{eq:Ediss}) then yields
[we use eq.~\ref{eq:gam_x} for $\gamma$ and replace $1+\beta$ by the 
step function $1+ H(1-x)$]
\begin{equation}
\label{eq:Ed}
  \Ed(x)\approx \Gej^2\macc c^2 
  \left\{\begin{array}{ll}
    x^k-\frac{6-k}{6+k}+\psi, & x>1, \\
    \frac{2kx^{6+k}}{6+k}+\psi, & \frac{1}{\sqrt{3}}<x<1, \\
    \frac{2k(x^{3+k}-\xgap^{3+k})}{3\sqrt{3}(3+k)}, & 
                                      \xgap<x<\frac{1}{\sqrt{3}},\\
  \end{array}\right.
\end{equation}
where $\psi\approx 0.004$ for $k=1$ and $\psi\approx 0.002$ for $k=3$.
Equation~(\ref{eq:Ed}) assumes a deceleration radius $\xdec>1$. 
Setting $\Ed=\Eej$ we find the actual deceleration radius,
\begin{eqnarray}
\label{eq:dec}
  \xdec\approx 
  \left\{\begin{array}{ll}
    \left[1+\frac{2k}{6+k}\left(\frac{1}{D}-1\right)\right]^{1/k}, & D<1,\\
    D^{-1/(6+k)}, & 1<D<27(\sqrt{3})^k,\\
  \end{array}\right.\\
\label{eq:D}
  D\equiv\frac{2k\Gej^2\macc c^2}{(6+k)\Eej}. \hspace*{3cm}
\end{eqnarray}
Note that $\xdec$ depends very weakly on $D$ at $D>1$. 
Equation~(\ref{eq:gam_x}) gives $\gamma(\xdec)$,
\begin{equation}
\label{eq:g_dec}
  \gdec=
  \left\{\begin{array}{ll}
    1, & D<1, \\
    D^{6/(6+k)}, & 1<D<27(\sqrt{3})^k.
  \end{array}\right.
\end{equation}
The simple formula~(\ref{eq:dec}) neglects $\psi$ and applies if
$D<27(\sqrt{3})^k$ (corresponding to $\xdec>1/\sqrt{3}$).
The extension to $D>27(\sqrt{3})^k$ is straightforward.
A much higher $D$, however, would imply that the swept medium is optically
thick, making the model inconsistent (we assumed transparency when building
the front model in \S\S~2 and 3). Let us evaluate the optical depth of the 
swept medium at $\Rgap$. The $e^\pm$ loading factor is 
$n^*/n_0\approx 2\times 10^2\mu_e/\xgap^2$ (cf.~eq.~\ref{eq:appr2}), and 
\begin{equation}
\label{eq:taugap}
 \taugap^*\approx\frac{2\times 10^2}{\xgap^2}\frac{\mgap\sT}{4\pi\Rgap^2m_p}
            =\frac{2\times 10^2}{\xgap^{4-k}}\frac{\macc\sT}{4\pi\Racc^2m_p}.
\end{equation}
The condition $\taugap^*<1$ defines a maximum $D_*$. Substituting 
$\xgap=\sqrt{3}\Gej^{-1/3}$ and using equations~(\ref{eq:Racc}) and 
(\ref{eq:D}) we find
\begin{equation}
\label{eq:Dstar}
  D_*\approx 0.1\left(\frac{\Eej}{E}\right)\Gej^{(2+k)/3}.
\end{equation}
With a typical $\Gej=300$, $D_*\approx 30$ for $k=1$ and $D_*\approx 10^3$
for $k=3$. If $D>D_*$ the trailing photons of the radiation front will be 
trapped and advected by the medium until it becomes optically thin.
Our front model does not apply to this case.

It is instructive to compute the swept mass at $\Rdec$
(using eqs.~\ref{eq:dec} and \ref{eq:m_x}).
In the regime $D<1$ ($\xdec>1$, $\gdec=1$) the standard estimate holds
(e.g. Rees \& M\'esz\'aros 1992),
\begin{equation}
\label{eq:mdec_st}
   \mdec\approx\frac{\Eej}{\Gej^2 c^2}, \qquad D<1.
\end{equation}
In the regime $D>1$ ($\xdec<1$, $\gdec>1$) the deceleration occurs in a 
relativistically moving medium; then 
\begin{equation}
\label{eq:mdec}
  \mdec\approx\frac{(6+k)\Eej}{2k\Gej^2c^2}D^{6/(6+k)}
       =\left(\frac{6+k}{2k}\right)\frac{\gdec\Eej}{\Gej^2c^2}.
\end{equation}

\medskip

If the ambient medium is ISM with a constant density 
$n_0\sim 1-10^2$~cm$^{-3}$ we have $k=3$ and 
$\macc=2.4\times 10^{24}\mu_en_0E_{53}^{3/2}$g. The parameter $D$ is then 
given by
\begin{equation}
\label{eq:D_ISM}
 D\approx 1.4\times 10^{-4}\mu_e n_0E_{53}^{3/2}{\Eej}_{53}^{-1}{\Gej}_2^2.
\end{equation}
The typical $D<1$ and hence $\xdec>1$ and the standard estimate 
(\ref{eq:mdec_st}) applies.
Equation~(\ref{eq:Ed}) yields the energy fraction that is dissipated at 
$x<1$: $\facc\approx (2/3)\xdec^{-3}\approx D\ll 1$. 
The fraction dissipated in the static 
pair-loaded zone, $1<x<\xload\approx \sqrt{5}$ (eq.~\ref{eq:relation}), is 
$\fload\approx (\xload/\xdec)^3\approx 20D$.

If GRBs have massive progenitors (e.g. Woosley 1993)
their ambient medium is a wind from the progenitor. From 
a Wolf-Rayet progenitor one expects a wind with mass loss 
$\dM\sim 10^{-5}M_\odot$~yr$^{-1}$ and velocity $w\sim 10^3$~km~s$^{-1}$ 
(Chevalier \& Li 1999).
In the case of a red giant, the wind velocity is smaller, 
$w\sim 10$~km~s$^{-1}$, and then the ambient density is higher.
The wind medium is described by $k=1$ and 
$\macc=(\dM/w)\Racc=7\times 10^{28}\dM_{21}w_8^{-1}E_{53}^{1/2}$~g.
The typical $D$ is then comparable to or much larger than unity,
\begin{equation}
\label{eq:D_wind}
    D=\frac{2\dM c^2{\Gej}^2}{7w\Eej}\Racc
   \approx 1.8\dM_{21}w_8^{-1}E_{53}^{1/2}{\Eej}_{53}^{-1}{\Gej}_2^2.
\end{equation}

To study the wind case in more detail we solve numerically 
equations~(\ref{eq:dyn1},\ref{eq:dyn2},\ref{eq:gam_x},\ref{eq:m_x}). 
Figure~4 shows the results for $\Gej=200$, $E=\Eej=10^{53}$erg, and
the efficiency $\eta=1$. 
The dissipation rate peaks at $x\approx\xdec$. Besides, a small local
maximum appears at $x=1$ [it is understood from eq.~\ref{eq:Ediss}:
$(1+\beta)/\gamma$ has a maximum at $\beta=0.5$ i.e. at $x=1$].
If $D\geq 1$ then $\xdec$ is close to unity and
one can see the strong peak of energy dissipation at $x\sim 1$.
For $D=100$, 80\% of the blast wave energy is dissipated 
at $0.5<x<1$ and 99\% at $0.3<x<2$. Note that in this model 
$D>D_*\approx 20$ (eq.~\ref{eq:Dstar}) and the initial stage $0.3<x<0.5$
is optically thick. 
Yet, since most of the energy is dissipated when the medium becomes
transparent, the model with $D=100$ is marginally applicable.

\centerline{ \epsfxsize=8.7cm {\epsfbox{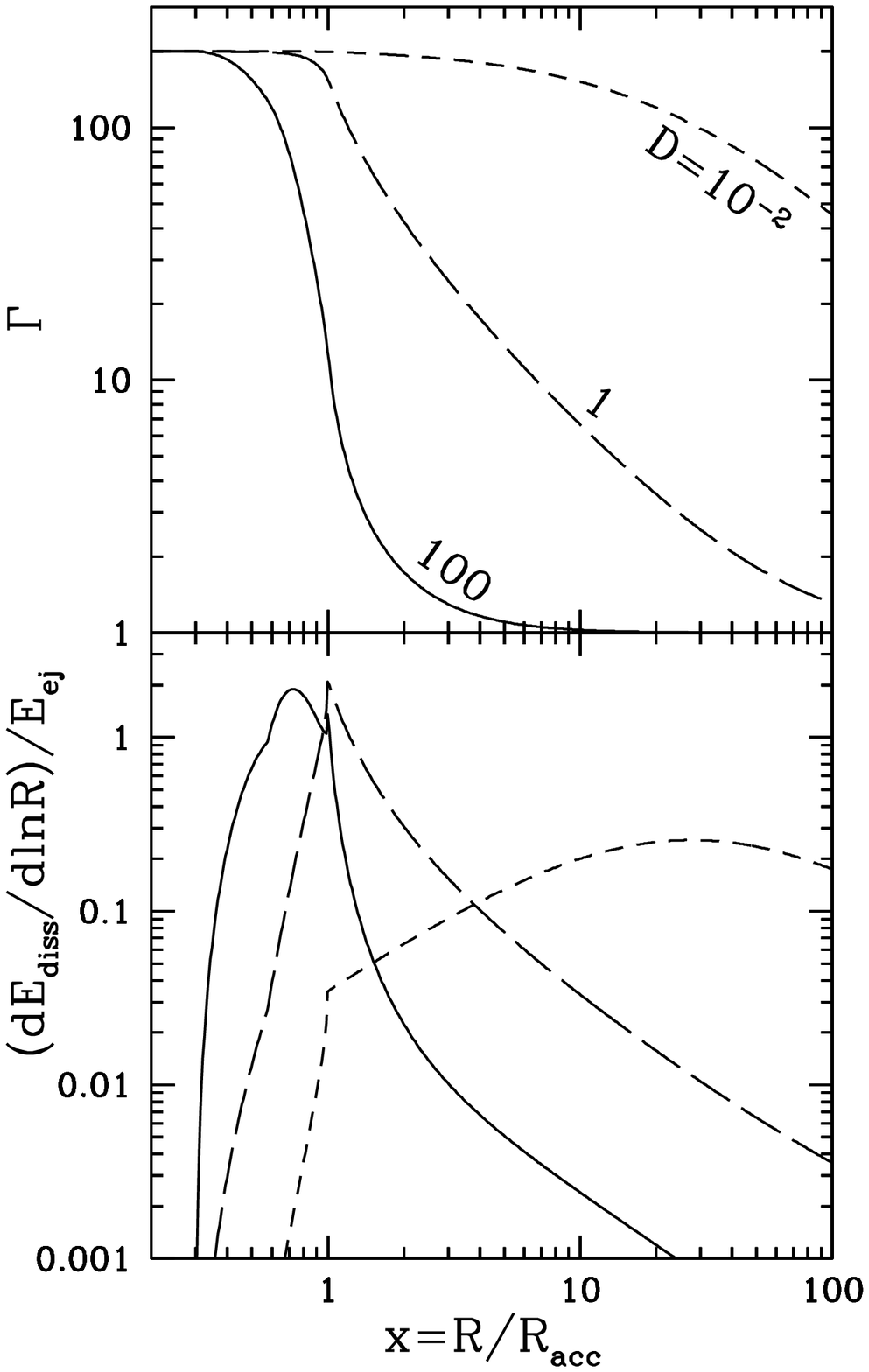}} }
\figcaption{ Blast wave dynamics in a wind. Here $E=\Eej=10^{53}$~erg,
$\Gej=200$, $\eta=1$. {\it Dashed, long-dashed, solid
curves} display the cases $D=10^{-2}$, 1, 100 (eq.~\ref{eq:D_wind}).
}
\bigskip

The dissipated energy $\Ed$ (all emitted if $\eta=1$) exceeds 
$\Eej$ because the additional energy $\Eacc=c^2\int (\gamma-1)\dd m$ is  
deposited by the radiation front when it preaccelerates the ambient medium. 
Using equation~(\ref{eq:gam_x}) one finds for a wind medium ($k=1$)
\begin{equation}
  \Eacc=\macc c^2\left(\frac{3\sqrt{3}}{2\xgap^2}-5.9\right)
       =\macc c^2\left(\frac{\sqrt{3}}{2}\Gej^{2/3}-5.9\right).
\end{equation}
The main contribution to $\Eacc$ comes from small radii $x\sim\xgap$.
Using equation~(\ref{eq:D}) one can find $\Eacc/\Eej$ as a function of $D$.
For example $\Eacc/\Eej\approx 2.1\times 10^{-3}D$ for $\Gej=200$.

The parameters $\eta=1$ and $\Eej=E$ taken in the numerical examples imply 
that the blast emits energy equal to that of the prompt GRB. This emission 
should contribute to the preacceleration (it is soft and 
preaccelerates efficiently, with no Klein-Nishina reduction). 
Here it was not accounted for; in the model with $D=100$ 
it would increase $\Racc$ by a factor of 2.

\subsection{Emission}

We now evaluate the main characteristics of the blast wave emission,
in particular, the bolometric light curve seen by a distant observer and the 
synchrotron peak frequency. A detailed analysis is deferred to a next paper.

\subsubsection{Bolometric light curve}

A distant observer will see a mixture of radiation emitted by the 
decelerating shell at 
different radii. Denote the arrival time of radiation by $\tobs$ and choose 
$\tobs=0$ for a light signal that would come from the center/beginning of 
the explosion. First consider the observed light curve from instantaneous
emission of energy $E_0$ by the shell at radius $R$. The shell reaches this
radius at a time $t(R)$ after the beginning of the explosion. Choose a polar 
axis $\theta=0$ pointing toward the observer. The observer 
will first receive photons emitted at $\theta=0$ ($\mu=\cos\theta=1$).
These first photons come at $\tobs=t(R)-R/c$ and photons emitted from 
a circle $\mu=const<1$ arrive with a delay of $(R/c)(1-\mu)$. We thus
have a relation 
\begin{equation}
\label{eq:time}
   \tobs(R,\mu)=t(R)-\frac{R}{c}\mu, \qquad 
  t(R)=\int_0^R\frac{\dd R}{\hat\beta c}.
\end{equation}
Radiation received in time interval $\delta\tobs$ comes from the ring 
$|\delta\mu|=\delta\tobs(c/R)$. The total energy emitted by this 
ring (in all directions) equals $\delta E=E_0|\delta\mu|/2$ where 
$|\delta\mu|/2$ is the fraction of the shell surface 
occupied by the ring. We will assume that each 
element of the ring emits isotropically in its rest frame. Radiation emitted 
toward the observer within a rest-frame solid angle $\dd\tilde\Omega$
occupies $\dd\Omega=\Gamma^2(1-\hat{\beta}\mu)^2\dd\tilde{\Omega}$ in the 
lab frame. Hence the observed flux is affected by the beaming factor 
$\Gamma^{-2}(1-\hat{\beta}\mu)^{-2}$ and the {\it apparent} isotropic 
energy seen by the observer from a ring $\delta\mu$ is 
$\delta E_{\rm app}=\Gamma^{-2}(1-\hat{\beta}\mu)^{-2}\delta E$.
(Integration of $\delta E_{\rm app}$ over the shell gives $E_0$ as it
should be.) The apparent isotropic luminosity is 
\begin{equation}
\label{eq:lum}
 \Lobs=\frac{\delta E_{\rm app}}{\delta\tobs}
  =\frac{cE_0}{2\Gamma^2R}\left\{1-\frac{\hat{\beta}c}{R}
     \left[t(R)-\tobs)\right]^2
                          \right\}^{-2}.
\end{equation}

From the dynamic solution we know $\dd\Ed/\dd R$ and $\Gamma(R)$.
It allows us to compute the observed light curve from the whole 
history of the shell deceleration [we substitute $E_0=\eta(\dd\Ed/\dd R)\dd R$
into eq.~\ref{eq:lum} and integrate over $R$],
\begin{equation}
\label{eq:Lobs}
  \Lobs(\tobs)=\int_0^{\Rmax}\frac{(R/c)\eta(\dd\Ed/\dd R)\dd R}
     {2\Gamma^2[R/c-\bh(t-\tobs)]^2},
\end{equation}
where $\Rmax(\tobs)$ is defined by condition $t(R)-R/c=\tobs$ 
(see eq.~\ref{eq:time}).

The results are shown in Figure~5 for the blast waves in wind environment
(same cases as in Fig.~4).  There is no emission until 
$\trise=\Rgap/2\Gej^2\approx E_{53}^{1/2}(\Gej/200)^{-7/3}$~s which 
corresponds to the moment when the ejecta catches up with the surfing 
medium and begins to decelerate. At $\tobs=\trise$ the light curve rises
steeply and then reaches a peak at 
\begin{equation}
\label{eq:tpeak}
  \tpeak\approx\frac{\Racc}{2\Gej^2} 
  \left\{\begin{array}{ll}
    1, & D<1 \\
    \xdec, & D>1
  \end{array}\right\}
 \approx 12\,\frac{E_{53}^{1/2}}{{\Gej}_2^2}{\rm~s}.
\end{equation}
Since $\xdec$ remains close to unity even at $D\gg 1$, we get a universal
$\tpeak$ in a very wide range of $D$. 
In the examples shown in Figure~5 ($E_{53}=1$, ${\Gej}_2=2$) we get 
$\tpeak\sim 3$~s. 
At $D<10^{-2}$ there appears a plateau in the light curve between $\tpeak$ 
and $\sim 0.3\Rdec/2\Gej^2c$. In the model with $D=100$ the total 
emitted energy exceeds $\Eej$ by 20\% owing to $\Eacc$ (\S~8.1).  

\centerline{ \epsfxsize=8.8cm {\epsfbox{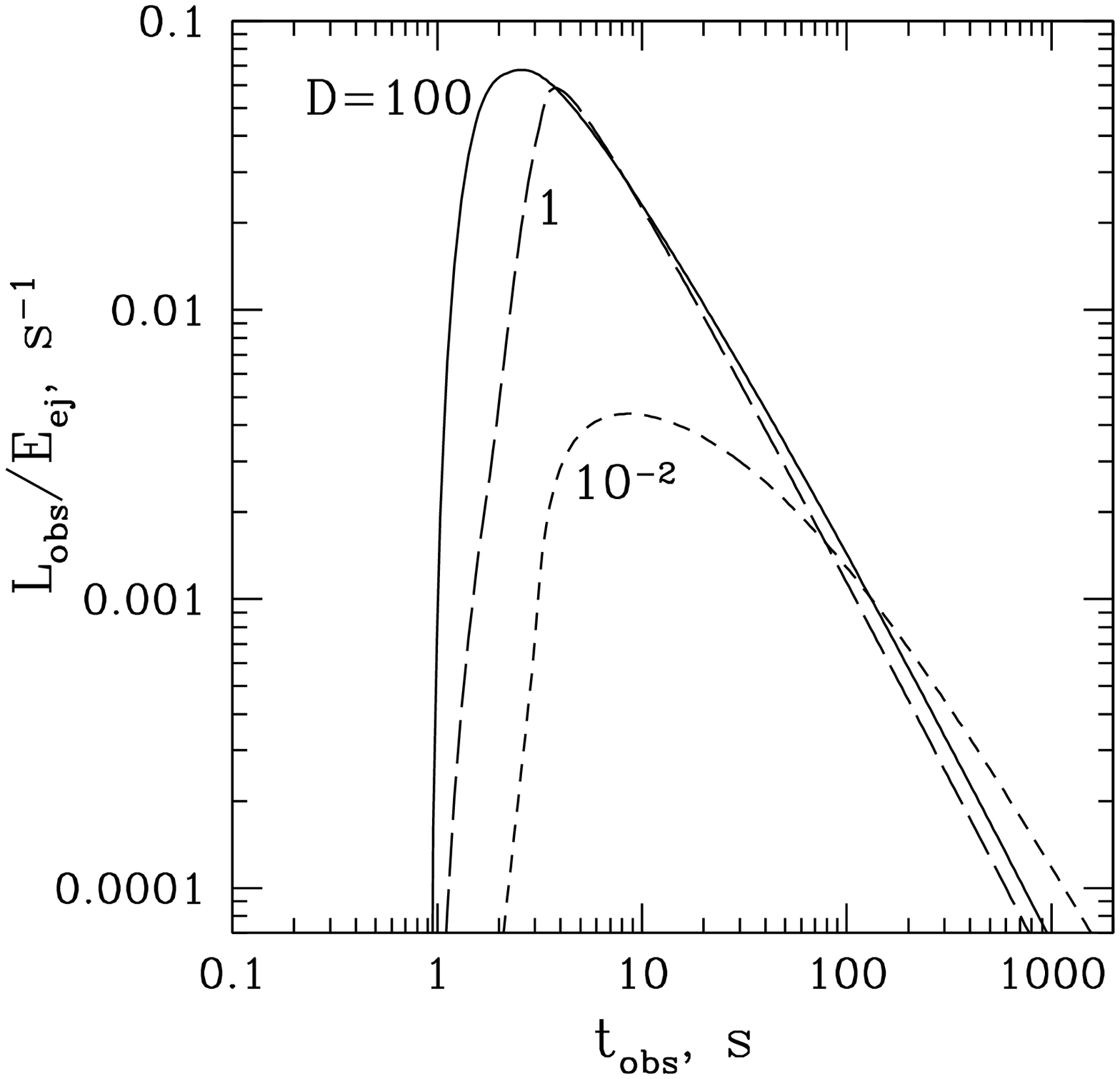}} }
\figcaption{ Bolometric light curves from the blast waves
shown in Figure~4.
}
\bigskip

Figure~5 assumes a radiative blast wave ($\eta=1$).
Adiabatic blast waves ($\eta\ll 1$) produce similar
light curves with the same $\tpeak$. In reality the radiative efficiency
is likely to change during the afterglow, being highest at $R\simlt\Racc$
(when a lot of $e^\pm$ are loaded) and decreasing at larger $R$. 
Then the peak will be sharper.

The light curve should further be corrected for the cosmological 
effects. In particular, for a burst with a redshift $z$ one gets
$\tpeak\approx 12(1+z)E_{53}^{1/2}{\Gej}_2^{-2}$~s.

\subsubsection{Synchrotron peak frequency}

The medium encountered by the blast wave is $e^\pm$-loaded, preaccelerated, 
and compressed by the leading radiation front. Correspondingly, 
the standard analysis of the blast wave emission (Blandford \& McKee 1977;
Piran 1999) applies to our case with three
modifications: (1) $e^\pm$ loading increases the number of shocked electrons 
by the factor $n^*/n_0$. (2) The proper mass density of the medium is 
$\tilde{\rho}=\rho_0\gamma^{-1}(1-\beta)$ rather than $\rho_0$; here the 
factor $(1-\beta)$ comes from compression in the front (cf.~\S~3.2) and 
$\gamma^{-1}$ appears due to Lorentz stretching when we go to the rest frame.
(3) The Lorentz factor of the pre-shock medium in the shock frame is 
$\Gamma/\gamma$ rather than~$\Gamma$.

The proper energy density of the post-shock material is 
\begin{equation}
\label{eq:u}
  u=4\left(\frac{\Gamma}{\gamma}\right)^2\tilde{\rho}c^2.
\end{equation}
Assuming that magnetic energy $B^2/8\pi$ is a fraction $\ep_B$ of $u$,
we have
\begin{equation}
 B=c\frac{\Gamma}{\gamma}\sqrt{\frac{32\pi\ep_B\rho_0}{\gamma(1-\beta)}}.
\end{equation}
Assuming that $e^\pm$ share the energy of shocked ions, the mean randomized 
Lorentz factor of $e^\pm$ in the rest-frame of shocked matter is
\begin{equation}
  \gamma_e=\frac{m_*}{m_e} \frac{\Gamma}{\gamma},
\end{equation}
where $m_*=\mu_em_pn_0/n^*$ and $1<\mu_e<2$ (eqs.~\ref{eq:ion} and \ref{eq:m}).

We then get the peak synchrotron frequency in the rest-frame of the shell,
\begin{equation}
\label{eq:tnup}
  \tnup\approx 10^6B\gamma_e^2{\rm ~Hz}\approx
    \frac{10^{30}\rho_0^{1/2}}{\gamma^{5/2}}\left(\frac{n_0}{n^*}\right)^2
    \mu_e^2\ep_B^{1/2}{\Gamma}_2^3{\rm ~Hz}.
\end{equation}
The corresponding observed frequency is $\nup=\tnup\Gamma(1+z)^{-1}$ where
$z$ is the cosmological redshift of the burst. 

For example consider a blast wave with $D\ll 1$ in ISM. 
At radii $\Racc<R<\Rdec$ we have $\Gamma\approx \Gej$ and $\gamma\approx 1$.
The density $n^*$ is given by equation~(\ref{eq:stat}) with 
$\vp/\vpload=(\Rload/R)^2$. Then from equation~(\ref{eq:tnup}) we get
\begin{eqnarray}
\label{eq:tnup1}
 \tnup\approx \frac{5\times 10^{18}\mu_e^{5/2}\ep_B^{1/2}n_0^{1/2}{\Gej}_2^3}
                   {\exp(2\Rload^2/R^2)+\exp(-2\Rload^2/R^2)+2}
   {\rm ~Hz},\hspace*{3mm}\\ \Racc<R<\Rdec.
\nonumber
\end{eqnarray}
For instance, the observed emission from $R=\Racc$ has the peak frequency 
$\nup\approx 2\times 10^{16}(1+z)^{-1}\ep_B^{1/2}\mu_e^{1/2}n_0^{1/2}
{\Gej}_2^4$~Hz. The emission from $R<\Racc$ is even softer, however,
the luminosity is small from that region ($\Ed/\Eej\approx D$ at $\Racc$, 
see \S~8.1).

As a second example consider a blast wave in a wind with $D>1$ and evaluate
the peak frequency at the deceleration radius $\Rdec<\Racc$. We substitute 
equations~(\ref{eq:dec}) and (\ref{eq:g_dec}) into equation~(\ref{eq:tnup}) 
and use $n^*/n_0\approx 74\mu_ex^{-4}$ (cf. eq.~\ref{eq:appr2}); 
$\rho_0$ is expressed in terms of $D$ using 
$\rho_0=(\dM/4\pi\Racc^2\xdec^2w)$ and equations~(\ref{eq:D_wind}) and 
(\ref{eq:Racc}). Then we get
\begin{equation}
  \tnup\approx\frac{2\times 10^{16}}{D^{2.6}}\;
  {\Gej}_2^2\ep_B^{1/2}{\Eej}_{53}^{1/2}E_{53}^{-3/4} {\rm ~Hz}. 
\end{equation}
For instance, with $D=10$ and $\ep_B=0.1$ the observed 
$\nup=\tnup\Gamma(1+z)^{-1}$ is close to optical/UV and then a large 
fraction of the blast wave energy is emitted in the optical band.

\subsection{Comparison with the standard afterglow model}

The impact of the radiation front on the blast wave
is illustrated in Figure~6. We take the usually assumed parameters 
of a GRB with a Wolf-Rayet progenitor and compare the model
developed in this section with the standard model that neglects the radiation 
front and has $\gamma=1$ and no $e^\pm$ ahead of the blast wave. 
For a simple illustration we assume a short GRB with an impulsive ejection
(so that the sweeping shell approximation can be used, cf. \S~8.1).
The synchrotron peak frequency $\nu_s$ is evaluated with $\ep_B=0.1$ which 
is close to its maximum possible value. 

\centerline{ \epsfxsize=8.7cm {\epsfbox{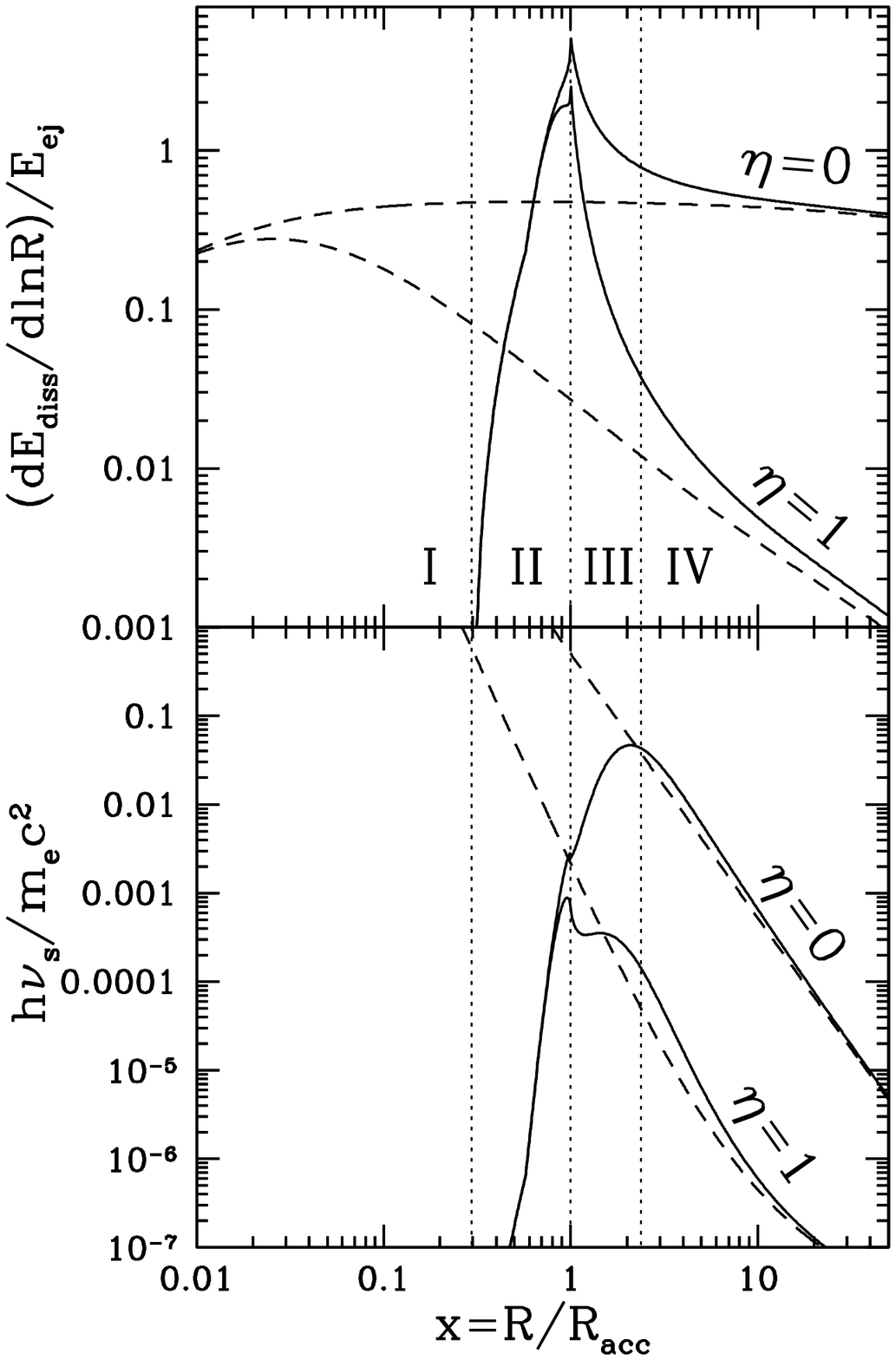}} }
\figcaption{ Afterglow from a GRB ejecta decelerating in a progenitor 
(Wolf-Rayet) wind with $\dM=2\times 10^{-5}M_\odot$~yr$^{-1}$, 
$w=10^{3}$~km~s$^{-1}$, and $\mu_e=2$.
The GRB is modeled as an impulsive emission of a gamma-ray front 
(with isotropic energy $E=10^{53}$~erg) and a thin ejecta shell 
with kinetic energy $\Eej=10^{53}$~erg and Lorentz factor $\Gej=200$. 
The assumed parameters imply $D\approx 10$ (eq.~\ref{eq:D_wind}).
{\it Dashed curves} show the prediction of the standard model that neglects
the impact of the radiation front and {\it solid curves} show the actual 
behavior. Two extreme cases are displayed in the figure: $\eta=0$ (adiabatic
blast wave) and $\eta=1$ (radiative blast wave). Four zones are marked:
I --- $R<\Rgap$ (the gap is opened),
II --- $\Rgap<R<\Racc$ (the gap is closed and the ejecta sweeps the 
relativistically preaccelerated $e^\pm$-loaded ambient medium), 
III --- $\Racc<R<\Rload$ ($e^\pm$-loaded ambient medium with 
$\gamma\approx 1$), and IV --- $R>\Rload$ (pair-free ambient medium 
with $\gamma\approx 1$). Radius is measured in units of
$\Racc=0.7\times 10^{16}E_{53}^{1/2}$~cm.
{\it Top panel}: the dissipation rate. {\it Bottom panel}: the synchrotron
peak frequency $\nu_s$ (assuming $\ep_B=0.1$) in units of $m_ec^2/h$. 
No cosmological redshift correction is applied here; the observed 
$\nu_s$ should be further reduced by $(1+z)^{-1}$. 
}
\bigskip

The front impact is dramatic in both radiatively efficient ($\eta=1$) 
and adiabatic ($\eta\ll 1$) cases: (1) no afterglow is emitted at $R<\Rgap$,
(2) the afterglow peaks sharply at $R=\Racc$, and (3) the main emission
occurs in the pair-loaded zones II and III and it is much softer than 
predicted by the standard model. Note that $\nu_s$ is highest at 
$R\approx\Racc$ for $\eta=1$ and at $R\approx\Rload$ for $\eta=0$. 
In the former case $\nu_s$ lies in soft bands at all $R$. For smaller 
$\ep_B$ and larger $D$ the afterglow will be even softer. One thus can 
expect ``X-ray-weak'' afterglows from GRBs with massive progenitors.

Note that the basic dynamic equations of \S~8.1 are not accurate in the 
adiabatic case. If not radiated, the dissipated energy is subject to 
adiabatic cooling and tries to accelerate the ejecta (or rather to reduce 
the deceleration rate). The present formulation neglects this effect and 
is in error by a factor of a few. Another special feature of the adiabatic 
regime is that the time-integrated dissipation rate far exceeds $\Eej$. 
The shell kinetic energy remains approximately constant and it is 
redissipated many times.


\section{Conclusions}

The unusual character of radiation fronts in GRBs is owing to
two basic facts: (1) the front is opaque for ambient electrons, i.e. 
the electron scatters many times when it is overtaken by the front, and 
(2) the front is opaque for the decollimated scattered photons 
($\gamma-\gamma$ opacity).
The two properties cause $e^\pm$ loading and violent acceleration 
of the medium.  The important role of the front is that it ``prepares'' 
the medium encountered by the blast wave. We summarize the medium dynamics
in the front in \S~9.1 and its impact on the ensuing blast wave in \S~9.2. 
The front should cause spectacular observational effects during 
the early afterglow. The expected phenomena are discussed~in~\S~9.3.

\subsection{The radiation front}

\begin{enumerate}

\item 
Photons scattered at one portion of the front get
absorbed at a different portion behind the location of scattering.
The local approximation used previously (assuming that the scattered photons
instantaneously become $e^\pm$) is not adequate: the
front structure is essentially governed by the radiative transfer.
Yet a simple
analytical description can be given to this non-local structure (\S~5).

\item
At radii $R>R_c$ (\S~7.2) the whole front has a
quasi-steady structure established on time-scales $\ll R/c$
(an ambient particle spends time $\delta t\ll R/c$ in the front).
The propagating front is described by a self-similar 
solution: the density amplification $n/n_0$ and the medium Lorentz factor
$\gamma$ are unique functions of  $\xi=\vp/\lambda$. 
Here $0<\vp<\Delta$ measures the distance inside the front ($\vp=0$ at the 
leading boundary) and $\lambda\propto R^2$ is the electron free-path
in the radiation field (eq.~\ref{eq:lam}). 
A simple analytical approximation to the structure functions is 
given in \S~5.
In the leading portion of the front 
the medium density exponentiates due to pair loading on length 
$\xiload\approx 30$, at $\xiacc\approx 5\xiload$ the medium is accelerated 
relativistically, and at $\xi_\pm\approx 30\xiload$ the loaded
pairs outnumber the ambient protons by the factor $m_p/m_e$ and dominate the 
inertia of the medium.
The medium parameters behind the front are $n(\xiD)$ and $\gamma(\xiD)$
where $\xiD=\Delta/\lambda\propto R^{-2}$ is its trailing boundary.
The front acts as a relativistic accelerator at radii 
$R<\Racc=7\times 10^{15}E_{53}^{1/2}$cm where $\xiD>\xiacc$. 
At $R>\Racc$ the front still loads the medium with $e^\pm$.
At $R>\Rload=1.5\times 10^{16}E_{53}^{1/2}$cm the $e^\pm$ loading is shut 
down ($\xiD<\xiload$).

\item
At radii $R<R_c$ the medium is accelerated so strongly that it 
gets ``stuck'' in the radiation front (the time-scale for the medium dynamics 
across $\Delta$ exceeds $R/c$). Then two zones exist in the front: 
(1)~$\xi<\xi_c\sim 10^3-10^4$ --- here the steady self-similar 
structure is established and (2) $\xi>\xi_c$ --- the ion-free zone.
Being strongly accelerated, the ambient medium cannot penetrate the 
zone $\xi_c<\xi<\xiD$.
Instead, it accumulates at $\xi\sim\xi_c$ and surfs the radiation pulse. 
With increasing $R$ the front traps new material which is accelerated to 
a smaller velocity. This causes the overshooting effect
and a caustic appears in the surfing medium (\S~7.2).
The surfing stage ends at the radius $R_c$ which is equal to 
or smaller than the radius of the blast wave formation.

\end{enumerate}

\subsection{The blast wave}

\begin{enumerate}

\item 
The blast wave forms at $R=\Rgap$. For short bursts
($t_b<4E_{53}^{1/2}{\Gej}_2^{-7/3}$s) 
$\Rgap\approx 3\times 10^{15}E_{53}^{1/2}{\Gej}_2^{-1/3}$cm and for longer 
bursts $\Rgap\approx 10^{16}t_b^{-1}E_{53}{\Gej}_2^{-8/3}$cm.
We limit our consideration here to short bursts.
Then the $\gamma$-ray front is detached from the blast wave (by a small
distance $R/2\Gamma^2$) and it leaves behind the changed ambient medium 
as described in \S~9.1 (item 2). In contrast,
radiation of long bursts continuously ``leaks out'' of the ejecta during
the blast wave stage which implies additional technicalities deferred to 
a future work.

\item
If the explosion happens in a constant-density medium with 
$\rho_0<2\times 10^{-22}{\Gej}_2^{-2}{\Eej}_{53}E_{53}^{-3/2}$~g~cm$^{-3}$ 
then the blast wave decelerates at $\Rdec>\Rload$ where the impact of the 
radiation front is small and the standard afterglow model works well. 
Such conditions probably take place for explosions in ISM.

\item
If the explosion happens in a wind from a massive progenitor 
the radiation front affects strongly the blast wave deceleration.
We defined a parameter $D$ (eq.~\ref{eq:D_wind}) that controls the dynamics;
the blast waves in winds have $D>1$
which corresponds to $\Rdec<\Racc$ i.e. the deceleration occurs 
in a relativistically moving medium. The standard estimate of $\Rdec$
is then invalid and instead one should use equation~(\ref{eq:dec}).
The ejecta decelerates near the unique radius 
$\Racc\approx 7\times 10^{15}E_{53}^{1/2}$~cm; it implies that $\Rdec$
weakly depends on the medium parameters as long as $D>1$. 
The ejecta does not decelerate at $R\ll\Racc$ because the medium ahead
has a high $\gamma$; at $R\simlt\Racc$ the medium preacceleration ceases
rather abruptly, $\gamma=(R/\Racc)^{-6}$, and the delayed deceleration 
occurs violently. 

\end{enumerate}

The strong effect of the radiation front on the blast wave formation and 
dynamics is easily understood. The front passes energy 
$\delta E\approx\delta\tau E/2$ to an ambient mass $\delta m$ ahead of 
the ejecta, where $\delta\tau\approx 0.2\sT\delta m (4\pi R^2m_*)^{-1}$ 
is the optical depth of $\delta m$ (here 0.2 is a Klein-Nishina 
correction and $m_*$ is mass per electron, $m_*<m_p$ due to pair loading). 
As a result the medium is accelerated to a Lorentz factor $\gamma$,
\begin{equation}
  \gamma-1=\frac{\delta E}{c^2\delta m}=\frac{\delta\tau}{\delta m}
   \frac{E}{c^2}
   =\frac{0.1\sT}{4\pi R^2 m_*}\frac{E}{c^2}.
\end{equation}
The front structure solution gives $m_*$ and $\gamma$ behind the front;
e.g. $m_*\approx m_p/74$ at $R=\Racc$ ($\beta=0.5$). 
At small radii, $R<\Rgap\sim\Racc/3$, one gets $\gamma>\Gej$, i.e.
the accelerated medium runs away from the ejecta and the gap is opened. 

The ejecta decelerates when it sweeps a sufficient mass $m$.
Namely, the deceleration condition reads $(\Gej/\gamma)m\approx\Mej$ i.e.
the swept inertial mass measured in the ejecta frame equals $\Mej$.
The radius of deceleration is increased if $\gamma\gg 1$ despite the fact 
that only a small energy $e\ll \Eej=\Gej\Mej c^2$ was used to preaccelerate 
the ambient medium. Indeed, we have [omitting the numerical factor 
$(6+k)/2k$, cf. eq.~\ref{eq:mdec}],
\begin{equation}
   \mdec\approx \frac{\gdec}{\Gej}\Mej, \quad 
   e=(\gdec-1)\mdec\approx \Eej\frac{\gdec(\gdec-1)}{\Gej^2}.
\end{equation}
The growth of $\mdec\propto\gdec$ occurs while $e<\Eej$ for $\gdec<\Gej$.

\subsection{Expected observational phenomena}

\begin{enumerate}
\item 
The generic prediction is that the early emission of a GRB blast wave
(at $t_{\rm obs}<30 E_{53}^{1/2}{\Gej}_2^{-2}$~s)
 should be very soft\footnote{
All the observed times given in this section are for GRB redshift $z=0$; 
they should be multiplied by $(1+z)$ if $z>0$.
}.
Compared to the standard model that neglects the effects of the radiation 
front,
the peak frequency of synchrotron emission is reduced by the pair loading
factor $(m_*/\mu_em_p)^2=(n^*/n_0)^{-2}$ and the preacceleration factor 
$\gamma^{-5/2}$ (see eq.~\ref{eq:tnup}).
The early afterglow 
should start as a relatively weak optical emission at $R<\Racc$ and
then the peak frequency moves to the X-ray band; only at $R>\Rload$ 
the blast wave sweeps the normal $e^\pm$-free medium with $\gamma\approx 1$ 
and emits in the standard regime.

\item
The fraction $f$ of the afterglow energy that is emitted at the early 
soft stage is controlled by the ratio $\Rload/\Rdec$. In the typical ISM 
environment $f<10$\%. In the typical wind environment (massive progenitor 
scenario) $f\approx 1$ i.e. most of the blast wave energy is emitted at 
the early soft stage. In the latter case we emphasize the likely 
possibility of ``X-ray-weak'' afterglows whose emission peaks in soft 
bands throughout the whole evolution of the blast wave (Fig.~6). 

\item
The light curves from blast waves in winds have special 
features which may be easily recognized in observations.
The violent deceleration that happens at $R\sim \Racc$ should cause 
a strong peak in the soft light curve. Then one should observe
(1) a steep rise of the afterglow at $\trise\approx \Rgap/2\Gej^2\approx 
E_{53}^{1/2}{\Gej}_2^{-2}$ and 
(2) a peak at $\tpeak\approx (\Racc/\Rgap)\trise\approx 
2.3{\Gej}_2^{1/3}\trise$.
Both $\trise$ and $\tpeak$ depend weakly on the wind parameters in a wide 
range $10^{-3}<D<10^2$ (\S~8.2.1).
Given the observed $E$ and $\tpeak$ one can find the Lorentz factor of
the ejecta. If $\Gej$ does not vary strongly from burst to burst
(as suggested by the clustering of GRB spectral peaks at $\ep\sim 1$, 
see Preece et al. 2000) there should exist a strong correlation between 
$\tpeak$ and the observed isotropic energy $E$ of the prompt GRB.

\item
In the massive progenitor scenario, the prompt high-energy $\gamma$-rays 
must be absorbed efficiently by radiation scattered in the wind (\S~6.2). 
As a result, the high-energy tail of the GRB spectrum will have a break.
The break position $\epbr$ is given by equation~(\ref{eq:break_wind}) for 
the idealized case of a burst with a constant flux; it appears at 
$\sim 10$~MeV in the beginning of the GRB and then slowly shifts with time
to higher energies. In highly variable 
bursts, future time-resolved spectroscopy should show a positive correlation
between $\epbr$ and the instantaneous flux.
Once the break is observed one can evaluate the wind density.

\end{enumerate}

The main observational effect of the radiation front is the strong softening 
of the early blast wave emission (which would otherwise be in the hard X-ray 
band). Owing to this softening the blast radiates in a different 
spectral window compared to the prompt GRB and it can be studied separately 
in simultaneous observations. Observations in optical -- soft X-ray bands 
at early times (less than $\sim 1$~min) can help to establish the nature of 
the GRB progenitor --- as we discussed here a wind
from a massive progenitor should have clear signatures.

Early optical emission has already been detected in GRB~990123 (Akerlof
et al. 1999) and it likely comes from the blast wave rather than 
internal dissipation in the ejecta since there is no correlation between 
the optical light curve and the prompt GRB. The optical emission can be
produced by the reverse shock in the ejecta (e.g. Sari \& Piran 1999). 
The results of the present paper suggest an
alternative interpretation: the soft emission is produced by the forward
shock at its early stage when it propagates in the 
preaccelerated and pair-loaded environment. 

The emission from the blast {\it hardens fast} when the blast wave
propagates in the $e^\pm$ loaded zone $R<\Rload$ and the observer will
see the whole spectrum from optical to X-ray bands. This broad-band emission
can overlap with the prompt GRB and its X-ray component can affect the 
measured GRB spectrum. In particular, the early external 
shock may generate the soft X-ray excesses detected in GRBs.

\medskip

The simple blast wave model constructed in this paper assumes a burst 
duration 
$t_b\leq t_b^*=4E_{53}^{1/2}{\Gej}_2^{-7/3}$s (cf. \S~7.2). After a
redshift correction $(1+z)\sim 3$, $t_b^*$ is still smaller than $\sim 10$~s, 
while a number of observed GRBs are longer than 10~s. The extension to 
$t_b>t_b^*$ should be simple since the found front structure in terms of 
$\xi$ (\S\S~4--6) will hold regardless $t_b$. However, the blast wave will 
be affected: (1) $\Rgap$ will decrease (eq.~\ref{eq:R_c_long}),
(2) $\xiD(R)$ [and hence $\gamma(R)$] will change, and
(3) the sweeping-shell approximation (\S~8) will not work; instead one
has to deal with the full hydrodynamical problem with the forward
and reverse shocks.


\acknowledgements

I thank A.F. Illarionov and C. Thompson for discussions and the referee,
R.D. Blandford, for comments. This work was supported by the Swedish Natural 
Science Research Council and RFBR grant 00-02-16135. 


\appendix

\begin{center}
  SCATTERING AND PHOTON-PHOTON ABSORPTION 
\end{center}

\subsection{Compton scattering}

The differential cross-section for Compton scattering (defined
in the electron rest frame) is given by (Jauch \& Rohrlich 1976)
\begin{equation}
  \frac{\dd\sigma}{\dd\tilde\mu}=\frac{3}{8}\sT
    \left(\frac{\tilde\epsc}{\tilde\ep}\right)^2\Psi, \qquad
  \Psi=\frac{\tilde\epsc}{\tilde\ep}+\frac{\tilde\ep}{\tilde\epsc}
            -2(1-\tilde\mu)+(1-\tilde\mu)^2, \qquad
  \frac{\tilde\epsc}{\tilde\ep}=\frac{1}{1-\tilde\ep(1-\tilde\mu)},
\end{equation}
where $\tilde\ep$ and $\tilde\epsc$
are the photon energies before and after scattering respectively 
(as measured in the electron rest frame), and $\tilde\mu$
is the cosine of the scattering angle in the rest frame. 
In our problem the scattering medium is cold and has a bulk velocity $\beta$ 
parallel to the direction of the primary collimated photons. The rest-frame 
magnitudes are then related to the lab ones by 
\begin{equation}
  \tilde\ep=\gamma(1-\beta)\ep, \qquad \tilde\epsc=\gamma(1-\beta\mu)\epsc,
  \qquad \tilde\mu=\frac{\mu-\beta}{1-\beta\mu},
  \qquad \frac{\dd\sigma}{\dd\mu}=\frac{\dd\tilde\mu}{\dd\mu}
                           \frac{\dd\sigma}{\dd\tilde\mu}
 =\frac{3}{8}\sT\left(\frac{\epsc}{\ep}\right)^2\frac{1+\beta}{1-\beta}\;\Psi.
\end{equation}
The Lorentz-invariant total cross-section is given by
\begin{equation}
 \sKN=\frac{3}{8}\frac{\sT}{\tilde{\ep}}\left[\left(1-\frac{2}{\tilde{\ep}}-
  \frac{2}{\tilde{\ep}^2}\right)\ln(1+2\tilde{\ep}) 
  +\frac{1}{2}+\frac{4}{\tilde{\ep}}-\frac{1}{2(1+2\tilde{\ep})^2}\right].
\end{equation}

\subsection{Saturation of radiative acceleration}

When the medium accelerates, the typical photon energy in the medium rest 
frame is redshifted well below $m_ec^2$ and the scattering occurs with Thomson 
cross-section, $\dd\sigma/\dd\tilde{\mu}=(3/8)\sT(1+\tilde{\mu}^2)$ and 
$\sKN=\sT$. With increasing $\gamma$ the finite collimation angle of the
radiation intensity $I(\theta)$ becomes important and the efficiency of 
radiative acceleration drops. The radiative force accelerating the electron 
is (Gurevich \& Rumyantsev 1965)
\begin{equation}
 \frac{\dd p}{\dd t}=\frac{\sT}{c}\,\gamma^2
  \left[I_1(1+\beta^2)-(I_0+I_2)\beta\right], 
 \qquad I_k=\int I(\theta)\,\cos^k\theta\,\dd\Omega. 
\end{equation}
Here $\theta$ is the angle between the ray and the radial direction.
Note that the net flux $F=I_1$.
If the radiation field is perfectly collimated ($I_0=I_1=I_2=F$) then 
$\dd p/\dd t=(\sT/c)F(1-\beta)/(1+\beta)$. For a finite collimation 
there exists a frame with a velocity $\bsat$ where the radiation flux 
vanishes. Assume that radiation is isotropic in this frame and has 
moments $\hat{I}_1=0$ and $\hat{I}_2=\hat{I}_0/3$, and compute $I_k$ in
the lab frame. The $I_k$ are
components of the stress-energy tensor of radiation, $I_0=cT^{00}$, 
$I_1=cT^{0x}$, $I_2=cT^{xx}$ (the $x$-axis is chosen along the radial 
direction). From the transformation law, 
$T^{ik}=\hat{T}^{lm}\Lambda^i_l\Lambda^k_m$ where $\bf{\Lambda}$ is the 
Lorentz matrix, one gets 
\begin{equation}
  I_0=\left(1+\frac{\bsat^2}{3}\right)\gsat^2\hat{I}_0, \qquad
  I_1=\frac{4}{3}\bsat\gsat^2\hat{I}_0, \qquad
  I_2=\left(\frac{1}{3}+\bsat^2\right)\gsat^2\hat{I}_0.
\end{equation}
Using these relations and substituting $F=I_1$ we find 
\begin{equation}
\label{eq:sat}
  \frac{\dd p}{\dd t}=\frac{\sT F}{c}\frac{\gamma^2}{\bsat}(\bsat-\beta)
  (1-\beta\bsat) \approx \frac{\sT F}{c}\left(\frac{1-\beta}{1+\beta}\right)
   \left[1-\frac{(1-\bsat)^2}{(1-\beta)^2}\right] 
  \approx \frac{\sT F}{c}\left(\frac{1-\beta}{1+\beta}\right)
   \left(1-\frac{\gamma^4}{\gsat^4}\right),
\end{equation}
where the approximate equalities make use of $\gsat\gg 1$.
The acceleration vanishes when $\gamma$ reaches $\gsat$.

\subsection{$\gamma-\gamma$ absorption}

The $\gamma-\gamma$ opacity seen by a scattered photon $(\mu,\epsc)$ is 
given by
\begin{equation}
\label{eq:opac}
  \kgg(\mu,\epsc)=\int_{\epthr}^{\epbr}
   \frac{F_\ep\sgg}{m_ec^3\ep}\;\dd\ep, \qquad 
  \epthr=\frac{2}{(1-\mu)\epsc},
\end{equation}
where $\sgg$ is the cross section for $\gamma-\gamma$
pair production (Jauch \& Rohrlich 1976)
\begin{equation}
\sgg(\ep_c)=\frac{3\sT}{8\ep_c^2}
 \left[\left(2+\frac{2}{\ep_c^2}-\frac{1}{\ep_c^4}\right)
       \ln\left(\ep_c+\sqrt{\ep_c^2-1}\right) 
      -\left(1+\frac{1} {\ep_c^2}\right)\sqrt{1-\frac{1}{\ep_c^2}}\;
 \right],
\end{equation}
and $\ep_c=(\ep/\epthr)^{1/2}$ is the energy of the
interacting photons in their center-of-momentum frame.
The mean energy of the photon absorbed by our photon $(\mu,\epsc)$ is
\begin{equation}
\label{eq:chi}
  \epabs(\mu,\epsc)=\int_1^{\sqrt{\epbr/\epthr}} \ep(\ep_c)P(\ep_c)\dd\ep_c
  =\chi\epthr, \qquad
   P(\ep_c)\dd\ep_c=\frac{F_\ep\sgg(\ep_c)\;\dd\ln\ep_c}
               {\int_1^{\sqrt{\epbr/\epthr}} F_\ep\sgg(\ep_c)\;\dd\ln\ep_c}.
\end{equation}
Here $P(\ep_c)$ is the probability of $\gamma-\gamma$ interaction with
given $\ep_c$ and $\ep(\ep_c)=\epthr\ep_c^2$. Thus defined numerical factor
$\chi$ depends on the spectrum shape $F_\ep$.
If the absorbing radiation has a power-law spectrum $F_\ep=F_1\ep^{-\alpha}$
one gets at $\epthr\ll\epbr$ 
\begin{equation}
\label{eq:opac_pw}
   \kgg=\frac{\ph(\alpha)}{\l1}\left(\frac{\epthr}{2}\right)^{-\alpha}, 
  \qquad \ph(\alpha)=2^{1-\alpha}\int_1^\infty\frac{\sgg}{\sT}
    \ep_c^{-2\alpha-1}\,\dd\ep_c,
  \qquad \chi=\frac{\ph(\alpha-1)}{2\ph(\alpha)},
\end{equation}
where $\l1=m_ec^3/F_1\sT$.
The numerical factor $\ph(\alpha)$ is with high accuracy ($<0.3$\% for
$0<\alpha<6$) approximated as
$\ph(\alpha)=(7/12)2^{-\alpha}(1+\alpha)^{-5/3}$ [Svensson 1987, eq.~B6 where
$\eta=2^{\alpha+1}(\alpha+2)^{-1}\ph$]. It gives $\chi=(1+\alpha^{-1})^{5/3}$.

The mean energy and momentum of the $e^\pm$ pair created when the
scattered photon gets absorbed are
\begin{equation}
\label{eq:gpm}
  e_\pm(\mu,\epsc)=\left(\epsc+\epabs\right)m_ec^2, \qquad 
  p_\pm(\mu,\epsc)=\left(\mu\epsc+\epabs\right)m_ec.
\end{equation}
In the rest frame of the medium, the mean Lorentz factor and momentum per 
injected particle are given by Lorentz transformation of the energy-momentum 
vector,
\begin{equation}
\label{eq:ginj}
 2m_ec^2\ginj(\mu,\epsc)=\gamma\left(e_\pm-\beta cp_\pm\right), \qquad
 2\pinj(\mu,\epsc)=\gamma\left(p_\pm-\beta \frac{e_\pm}{c}\right).
\end{equation}
It is straightforward to further average $\ginj$ and $\pinj$ over
the whole primary spectrum and scattering angles.

\subsection{The effective Klein-Nishina cutoff}

To the first approximation, $\epKN\sim \gamma$. This estimate is 
sufficient if $\epKN\gg 1$, far from the spectrum peak.
However at $\gamma\sim 1$ we have $\epKN$ near the peak and the results
of the analytical model in \S~5 are sensitive to the exact position of $\epKN$. 
The effective $\epKN$ depends on what we calculate. In calculations of 
$\dnp$ (eq.~\ref{eq:ndot} of the paper), $\dd\sigma$
enters in combination with opacity $\kgg$ seen by the scattered photon.
Let us assume $\beta=0$ and compute the average 
\begin{equation}
  \sKN\overline{\kgg}(\ep)=\int \dd\sigma \kgg(\mu,\epsc)
    =\frac{\sT}{\l1}\hat{\phi}(\alh)\ep^{\alh}X_{\alh}(\ep), \qquad
  X_{\alpha}\equiv\overline{\left[(1-\mu)\frac{\epsc}{\ep}\right]^{\alpha}}
    =\frac{\sKN}{\sT}\int_{-1}^{1}\dd\mu\,\frac{\dd\sigma}{\dd\mu}
     \left[(1-\mu)\frac{\epsc}{\ep}\right]^{\alpha}.
\end{equation}
Here bar denotes the averaging over scattering angles.
At $\ep\ll 1$ we are in the Thomson regime with 
$X_{\alpha}=\XT_\alpha=(3/8)\int (1-\mu)^{\alpha}(1+\mu^2)\dd\mu$, e.g. 
$\XT_2=7/5$. The Klein-Nishina correction is important ($\sim 1/2$) 
already at $\ep\sim 0.1$. It is due to two effects: 
(1) the scattering angle is reduced [and $\kgg\propto(1-\mu)^{\alh}$] and 
(2) the total cross-section $\sKN$ is reduced. 
Photons of energy $(\ep,\ep+\dd\ep)$ contribute to $\dnp$
with approximate weight $\propto (F_\ep/\ep)\sKN\overline{\kgg}$, 
therefore we define the effective $\epKN$ for $e^\pm$ loading as
\begin{equation}
  \int_0^\infty\dd\ep\frac{F_\ep}{\ep}\ep^{\alh}X_{\alh}
  =\int_0^{\epKN}\dd\ep\frac{F_\ep}{\ep}\ep^{\alh}\XT_{\alh}, \qquad
  \epKN=\left[(\alh-\als)\int_0^\infty\dd\ep 
   f_\ep\ep^{\alh-1}\frac{X_{\alh}}{\XT_{\alh}}\right]^{1/(\alh-\als)}
   \approx 0.4,
\end{equation}
where $f_\ep\equiv F_\ep/F_1$.
We get $\epKN\approx 0.4$ for $\als=0$ and $\alh=1.5$, 2, 2.5. 

When calculating $\Ppm$ (eq.~\ref{eq:Ppm} of the paper), we need to evaluate
\begin{equation}
  \sKN\overline{\kgg p_\pm}(\ep)=\int \dd\sigma \kgg(\mu,\epsc)p_\pm
   \approx\frac{\sT}{\l1}\hat{\phi}(\alh-1)\ep^{\alh-1}X_{\alh-1}(\ep),
\end{equation}
where we neglected the $\overline{\kgg\mu\epsc}$ term in 
$\overline{\kgg p_\pm}$.
Photons of energy $(\ep,\ep+\dd\ep)$ contribute to $\Ppm$
with approximate weight $\propto (F_\ep/\ep)\sKN\overline{\kgg p_\pm}$
(see eq.~\ref{eq:Ppm} of the paper), and the effective $\epKN$ is given by
\begin{equation}
  \int_0^\infty\dd\ep\frac{F_\ep}{\ep}\ep^{\alh-1}X_{\alh}
  =\int_0^{\epKN}\dd\ep\frac{F_\ep}{\ep}\ep^{\alh-1}\XT_{\alh-1}, \qquad
  \epKN=\left[(\alh-\als-1)\int_0^\infty\dd\ep f_\ep\ep^{\alh-2}
  \frac{X_{\alh-1}}{\XT_{\alh-1}}\right]^{\frac{1}{\alh-\als-1}}\approx 0.4.
\end{equation}
We assumed here that $\alh-\als>1$. 
Again we get $\epKN\approx 0.4$ for $\als=0$ and $\alh=1.5$, 2, 2.5.

In a similar way, one can evaluate the effective $\epKN$ for $\Psc$
(see eq.~\ref{eq:Psc} of the paper),
\begin{equation}
   \sKN\overline{(1-\mu\frac{\epsc}{\ep})}=\sT Z, \qquad
  \int_0^\infty\dd\ep F_\ep Z=\int_0^{\epKN}\dd\ep F_\ep Z^{\rm T}, \qquad
 \epKN=\int_0^\infty\dd\ep f_\ep\frac{Z}{Z^{\rm T}}\approx 0.7. 
\end{equation}
Here $Z^{\rm T}=(1+\beta)^{-1}=1$ at $\beta=0$.

When the medium accelerates, $\epKN$ increases well above unity and ends up 
outside the spectrum  peak; then its exact value is unimportant.
A sufficient approximation at $\beta\geq 0$ is 
$\epKN(\beta)=\epKN(0)\gamma(1+\beta)$ i.e. just the Doppler shifted value
found at $\beta=0$.



\begin{references}


\reference{}
Akerlof, C. et al. 1999, Nature, 398, 400

\reference{}
Blandford, R. D., \& McKee, C. F. 1977, MNRAS, 180, 343

\reference{}
Chevalier, R. A., \& Li, Z.-Y. 1999, ApJ, 520, L29

\reference{}
Dermer, C. D., \& B\"ottcher, M. 2000, ApJ, 534, L155

\reference{}
Gurevich, L. E., \& Rumyantsev, A. A., 1965, Sov. Physics -- JETP, 20, 1233

\reference{}
Jauch, J. M., \& Rohrlich, F. 1976, The Theory of Photons and Electrons,
Springer, New York

\reference{}
Madau, P., \& Thompson, C. 2000, ApJ, 534, 239

\reference{}
M\'esz\'aros, P., Ramirez-Ruiz, E., \& Rees, M. J. 2001, ApJ, 554, 660

\reference{}
Piran, T. 1999, Phys. Rep., 314, 575

\reference{}
Preece, R. D., Briggs, M. S., Mallozzi, R. S., Pendleton, G. N., 
Paciesas, W. S., \& Band, D. L., 2000, ApJS, 126, 19

\reference{}
Rees, M. J., \& M\'esz\'aros, P. 1992, MNRAS, 258, 41

\reference{}
Sari, R., \& Piran, T. 1999, ApJ, 517, L109

\reference{}
Smolsky, M. V., \& Usov, V. V. 2000, 531, 764

\reference{}
Svensson, R. 1987, MNRAS, 227, 403

\reference{}
Thompson, C., \& Madau, P. 2000, ApJ, 538, 105 (TM)

\reference{}
Woosley, S. E. 1993, ApJ, 405, 273

\end{references}
\end{document}